\documentclass{aa}
\usepackage{graphicx}
\begin{document}

\title{A unified  equation of state of dense matter  
and 
neutron star structure}
\author{F. Douchin \inst{1,3}
 \and
P. Haensel\inst{2}
}
\institute{Department of Physics, University of Illinois at Urbana-Champaign, 
           Urbana, Illinois 61801, U.S.A.\\
\and
N. Copernicus Astronomical Center, Polish
           Academy of Sciences, Bartycka 18, PL-00-716 Warszawa, Poland\\
\and
Centre de Recherche Astronomique de Lyon, ENS de Lyon, 69364 Lyon, France\\
  }
\offprints{P. Haensel, {\tt haensel@camk.edu.pl}}
\date{
 Received 31 July 2001/ Accepted 10 September 2001 
}
\abstract{
An equation of state (EOS) of neutron star matter, describing
both the neutron star crust and the liquid core, is calculated. 
It is based on the effective nuclear interaction  SLy of the 
Skyrme type, which is particularly suitable for the application 
to the calculation of the properties of very neutron rich matter 
(Chabanat et al. 1997, 1998). The structure of the crust, and its 
EOS, is calculated in the $T=0$ approximation, and under the assumption 
of the ground state composition. The crust-core 
transition is a very weakly first-order phase transition, with 
relative density jump of about one percent.  The EOS of the liquid core 
is calculated assuming  (minimal)  $npe\mu$ composition. 
Parameters of static neutron stars are calculated and compared with 
existing observational data on neutron stars.  
The minimum and 
maximum masses of static neutron stars are $0.094~{\rm M}_\odot$ and 
$2.05~{\rm M}_\odot$, respectively.  Effects of rotation on the minimum 
and the maximum mass of neutron stars are briefly discussed. 
\keywords{dense matter -- equation
 of state -- stars: neutron -- stars}
}

\titlerunning{ Equation of state of neutron star interior}
\authorrunning{F. Douchin and P. Haensel}
\maketitle

\section{Introduction}
%
The equation of state (EOS) of dense neutron star matter is 
one of the mysteries of these objects. The EOS is a 
basic input for construction of neutron star models. Its 
knowledge is needed  to calculate 
 the properties of neutron stars, 
which
in turn is   necessary for modelling many astronomical objects 
and phenomena. In particular, the knowledge of the EOS is 
necessary for the determination of the maximum mass of neutron stars, 
$M_{\rm max}$: compact objects  with $M>M_{\rm max}$ 
could not be but black holes. 

The EOS is predominantly determined by the nuclear (strong) 
interaction between elementary constituents of dense matter. 
Even in the neutron star crust, with density below normal 
nuclear density $\rho_0=2.7~10^{14}~{\rm g~cm^{-3}}$ (corresponding 
to baryon 
density $n_0=0.16~{\rm fm^{-3}}$), nuclear interactions 
are responsible for the 
properties (and actually - for the very existence!) 
of neutron rich nuclei, crucial for the crust EOS. The knowledge 
of these interactions is particularly important for the 
structure of the inner neutron star crust, where nuclei are 
immersed in a neutron gas, and even more so for the EOS 
of the liquid core. Nuclear interactions 
are actually responsible for a dramatic lifting of 
$M_{\rm max}$ 
from $0.7~{\rm M}_\odot$,  obtained when  
interactions are switched-off (Oppenheimer \& Volkoff 1939), 
above  measured $1.44~{\rm M}_\odot$ of PSR B1913+16, and 
maybe even above  
 $2~{\rm M}_\odot$, as suggested by some models of the kHz 
quasi periodic oscillations in 4U 1820-30 (Zhang et al. 1997, 
Miller et al. 1998, Klu{\'z}niak 1998).  

The outer envelope of neutron star with $\rho<\rho_0$ 
 contains the same elementary 
constituents as ordinary (e.g. terrestrial) matter, i.e., 
protons, neutrons, and electrons. 
Unfortunately, even within this  subnuclear density envelope,  
 calculation of the 
EOS starting from an experimentally determined 
{\it bare} nucleon-nucleon (NN) interaction {\it in vacuum},  
supplemented with a three-nucleon 
 (NNN) force (which is necessary to fit simultaneously 
with two body data also the properties of $^{3}{\rm H}$ and 
$^{4}{\rm He}$),  is not feasible. This is 
due to  the prohibitive 
complexity of the many-body problem to be solved in the case of 
 heavy  nuclei (more generally: for  nuclear structures - 
spheres, rods, plates etc. (Lorenz et al. 1993)) 
 immersed in a neutron gas. 
To make a calculation  feasible, one uses a mean field 
approximation  
with an {\it effective}  NN interaction, an approach used 
with great success in terrestrial nuclear physics. The most 
ambitious application of this approach to 
 the determination of the structure and EOS 
of the neutron star crust remains the classical work of Negele 
\& Vautherin (1973). Other authors, who treated this 
problem,  used additional  approximations 
of the quantum mean-field scheme 
(see:  Oyamatsu 1993, Lorenz et al. 1993, 
Sumiyoshi et al. 1995, Cheng et al. 1997, Douchin \& Haensel 
2000, and references therein). 

It is clear that in order to describe in a physically 
(in particular, thermodynamically)  
consistent way both the crust, the liquid core, and the transition 
between them, one has to use 
the same many-body model and the same effective 
NN interaction,  on both sides of the crust-core 
interface. The mean-field scheme can also be applied for 
the description of the spatially uniform $npe$ liquid 
provided one uses appropriate effective 
NN interaction. Notice, that in this case the 
calculation of the ground state of nucleon matter 
can be done also, with rather high precision (at not too high 
density),  starting with 
bare nuclear hamiltonian ${\hat H}_{\rm N}$ (resulting 
from bare NN and NNN interactions) (Wiringa et al.1988, 
Akmal et al. 1998). The calculated ground state energy  
$\langle \Psi_0\mid {\hat H}_{\rm N}\mid \Psi_0\rangle$ (here 
$\Psi_0$ is the actual ground state wave function, which includes 
nucleon correlations, and minimizes the energy of the system) 
has then to be 
approximated, as well as possible, by  
$\langle \Phi_0\mid {\hat H}_{\rm N}^{\rm eff}\mid\Phi_0\rangle$,
where $\Phi_0$ is the Hartree-Fock wave function, 
and ${\hat H}_{\rm N}^{\rm eff}$ is effective nuclear 
hamiltonian.

Some authors formulated the nuclear many-body problem, 
relevant for neutron star matter, within  relativistic 
mean-field models, in which nuclear interactions 
are described by a phenomenological lagrangian involving
coupling of the nucleon fields to the meson fields (Sumiyoshi 
et al. 1995, Cheng et al. 1997). While such an approach 
has an obvious advantage at very high density (it yields 
causal EOS, by construction), its meaning at lower 
densities is not clear (see, e.g., Heiselberg \& Pandharipande 
2000). In the present paper we restrict ourselves to the 
non-relativistic approach.

Once the many-body approximation was fixed, the input 
consists of the effective NN interaction, which has to 
reproduce  a wealth of experimental data on atomic nuclei, 
especially those with high neutron excess,  
as well as reproduce  the most 
reliable numerical results concerning the ground state of 
dense homogeneous neutron rich nucleon matter. In the case 
of the calculation of the EOS of neutron star matter, 
the latter 
condition may be  reduced to the limiting 
case of pure neutron matter; as it turns out, in such a case 
 many-body calculations with bare nucleon hamiltonian 
are particularly precise, 
mostly because of the less important role played by 
 the tensor forces (Wiringa et al. 1988).

An effective nucleon hamiltonian 
contains a number of parameters which are usually fixed 
by fitting experimental data on saturation properties of 
bulk nuclear matter and experimental properties of 
selected atomic nuclei. The parameters of 
${\hat H}_{\rm N}^{\rm eff}$ 
are also constrained by 
some general condition, e.g., of spin stability (Kutschera \& 
W{\'o}jcik 1994). 
Most of the  existing effective 
interactions were  fitted to the properties of laboratory atomic
nuclei, with $(N-Z)/A < 0.3$,
 while in the bottom layers of neutron-star crust, and 
even more so in the liquid core,  one expects
$(n_n-n_p)/n_{\rm b}\ga 0.8$. In view of this, application of
these effective nuclear interactions to neutron star interior
 involves a rather risky extrapolation to
strongly asymmetric nucleon matter. In order to remove a part of
this uncertainty, modifications of  effective nuclear  forces, to make
them consistent with available (and possibly reliable) results of 
microscopic calculations of neutron matter,
have been applied. Such a procedure was used in  seventies to
obtain the Sk$1^\prime$ force 
(Lattimer \& Ravenhall 1978), via a rather {\it
ad hoc} modification of the Sk1 force constructed originally by
Vautherin and Brink 
(Vautherin \& Brink 1970) to describe terrestrial
nuclei. In this way, Sk$1^\prime$  
became consistent with energy per nucleon
of neutron matter calculated by Siemens and Pandharipande (1971). 
 Later, generalized types of the Skyrme
interaction, FPS (Pandharipande \& Ravenhall 1989) 
 and FPS21 (Pethick et al. 1995),
with a larger number of fitted parameters and more general density
dependence, were derived by fitting  the 
temperature and baryon  density
dependent energies per baryon of nuclear and neutron matter
obtained in microscopic calculations of Friedman and Pandharipande
(1981). 

A new  set of the Skyrme-type effective N-N interaction has
been derived recently, based on an approach which may be more appropriate, as
far as the applications to a very neutron rich matter are concerned
(Chabanat et al. 1997, Chabanat et al. 1998). While being of a two-body 
type, this effective interaction  contains, 
in the  spirit of the Skyrme model, 
 a  term resulting  from averaging 
of an original  three-body component. 
 Relevant additional experimental items
concerning neutron rich nuclei (including isovector effective
masses), constraints of spin stability and requirement of consistency
with the UV14+VIII equation of state (EOS)
  of dense neutron matter of Wiringa et al. (1988)
  for $n_0\le n_{\rm b}\le 1.5~{\rm fm^{-3}}$ were combined with
the general procedure of fitting the properties of doubly
 magic nuclei. This procedure led to a set of the 
SLy ({\bf S}kyrme {\bf Ly}on) effective nucleon-nucleon 
interactions
 which - due to the emphasis
 put on their neutron-excess dependence - seem to be particularly
suitable for the calculations of the properties of neutron-star
interiors. 
The FPS force was constructed as a generalized Skyrme model, 
by fitting the properties of asymmetric dense, cold and hot, 
nucleon matter, calculated by Friedman and Pandhripande (1981); 
fitting of the  ground state properties of laboratory nuclei 
was  not included in their derivation. Luckily, the FPS force 
turned out to reproduce rather well, without 
additional adjustement,  the ground state
energies of eight doubly closed-shell nuclei ranging from 
$^{16}{\rm O}$ to $^{208}{\rm Pb}$ (Lorenz et al. 1993).  

The SLy forces have been constructed as to be consistent with the
UV14+UVII  model of Wiringa et al. (1988) of neutron matter
{\it above} $n_0$ (Chabanat et al. 1997, Chabanat et al. 1998). 
It is therefore of interest to check  how well these effective N-N
interactions reproduce 
 the  UV14+UVII equation of state  of neutron matter at
subnuclear densities. This feature is quite important for 
the correct calculation of the equation of state of the 
bottom layers of neutron star crust and of the 
liquid core, which contain only a few 
percent of protons. We do not have direct access to the ``experimental  
equation of state'' of pure neutron matter at subnuclear densities. 
However, results of the best numerical many-body calculations 
of the ground state of neutron matter with realistic 
${\hat H}_{\rm N}$ seems to be sufficiently precise at subnuclear 
 densities to be used as an {\it ersatz} of experimental data 
(Pethick et al. 1995). The SLy effective interaction passes 
this test very well, in contrast to most of other models of
${\hat H}_{\rm N}^{\rm eff}$ 
 (Douchin \& Haensel 2000). 
In what follows, by the SLy interaction we will mean the basic 
SLy4 model of Chabanat et al. (1998). 

 After our unified EOS was constructed, a new state-of-the-art 
microscopic calculation of the EOS of dense matter (Akmal et al. 1998),  
 which in many 
respect is superior to the ten years older   models of Wiringa et al. (1988),
became available. The nuclear hamiltonian of Akmal et al. (1998) is based 
on a new Argonne two-nucleon interaction  AV18, takes into account relativistic 
boost corrections to the two-nucleon interaction, and includes new Urbana 
model of three-nucleon interaction, UIX. In what follows, the most complete 
models of the EOS of dense cold catalyzed matter, ${\rm AV18+\delta v + 
UIX^*}$, calculated by Akmal et al. (1998), will be referred to as APR 
({\bf A}kmal {\bf P}andharipande {\bf R}avenhall). It should be stressed, that 
in contrast to the FPS and SLy EOS, the APR EOS describes only the liquid core 
of neutron star, and therefore is not  a ``unified EOS'' of the neutron star 
interior. As we will show, neutron star models based on the APR EOS of the 
liquid core,  supplemented 
with our EOS of the crust, are not very different from the stellar
 models calculated 
using our complete, unified EOS.

In the present paper we calculate the unified EOS for neutron 
star matter using the SLy effective NN interaction. 
Nuclei 
in the crust are decribed using the Compressible Liquid 
Drop Model, with parameters calculated using the 
many-body methods 
presented in 
Douchin et al. (2000) and Douchin \& Haensel (2000). 
The calculation of the EOS is 
continued to higher densities, characteristic of 
the liquid core 
of neutron star. Using our EOS, we then calculate  
 neutron star models and compare their 
parameters with those 
obtained using older FPS effective NN interaction. 
We consider also effects
of rotation on neutron star structure. 
Our neutron star models are  then confronted 
with observations of neutron stars. 

The method of the calculation of the EOS for the crust 
and the liquid core of 
neutron star is described in Sect.2.  
Results for the structure and the EOS of the crust are given 
in Sect.3, and those for the liquid core in Sect.4. 
Models of neutron stars are reviewed  in Sect.6. 
Effects of rotation on neutron star structure 
are briefly discussed in the two last subsections 
of Sect. 6. Comparison with observations 
of neutron stars is presented in Sect.7. Finally, 
Sect.8 contains summary and conclusion of our paper.  

\section{Method of solution of the many-body problem}
\subsection{The crust}
%
Nuclei in the  neutron-star crust are  
described using the Compressible Liquid Drop Model (CLDM) of
nuclei 
(Douchin et al. 2000 and references therein). 
The parameters of the model are calculated 
using the SLy effective interaction. 
 Within the
CLDM, one is able to separate bulk, surface and Coulomb
contributions to total energy density, $E$. Electrons are assumed
to form a uniform Fermi gas and yield the rest plus  kinetic
energy contribution, denoted by $E_e$. Total energy
density of the neutron-star crust is given by
\begin{equation}
E=E_{{\rm N,bulk}} + E_{{\rm N,surf}} + E_{\rm Coul} + E_{e}~.
\label{E.CLDM}
\end{equation}
Here, $E_{{\rm N,bulk}}$ is the bulk contribution of nucleons,
which does not depend on the shape of nuclei. However,
both $E_{{\rm N,surf}}$ and  $E_{\rm Coul}$ do depend on this 
 shape. 
 Actually, in the case of the bottom layers of the 
inner crust, one has to generalize the notion of ``nuclei'' 
to ``nuclear structures'' formed by denser nuclear matter and  the
less dense neutron gas. Detailed description of the calculation of
$E_{\rm N,surf}$ and $E_{\rm Coul}$ for the SLy forces was
presented in Douchin (1999)and Douchin et al. (2000). 
Its application to the calculation of the properties of the 
inner crust is presented in Douchin \& Haensel (2000).  
 We restricted ourselves to three
shapes of the nuclear matter - neutron gas interface: spherical,
cylindrical, and plane. Consequently, we considered five types of
nuclear structures: spheres of nuclear matter in neutron gas,
cylinders of nuclear matter in neutron gas (rods), plane slabs of
nuclear matter in neutron gas, cylindrical holes in nuclear matter
filled by neutron gas (tubes) and spherical holes in nuclear
matter filled by neutron gas (bubbles). 
In view of a significant
neutron excess, the interface includes neutron skin formed by
neutrons adsorbed onto the nuclear matter surface. In view of a
finite thickness of nuclear surface, the definition of its spatial
location is a matter of convention. Here, we defined it by the
radius of the equivalent constant density  proton 
distribution, $R_p$, which
determines thus the radius of spheres, bubbles, cylinders and
tubes, and the half-thickness of plane nuclear matter slabs. 
The neutron radius, $R_n$, 
was defined by the condition that it yields a squared-off neutron 
density distribution with constant neutron densities, which are 
equal to the real ones far from the nuclear matter -- neutron gas 
interface, 
and reproduces actual total number of neutrons. The thickness of the 
neutron skin was  then defined as  $R_n-R_p$. 
The
nuclear surface energy term, $E_{{\rm N,surf}}$, gives the
contribution of the interface between  neutron gas and  nuclear
matter; it includes contribution of neutron skin
(Pethick \& Ravenhall 1995, Lorenz 1991). 
 In the case of spherical and
cylindrical interface, $E_{{\rm N,surf}}$ includes curvature
correction; the curvature correction vanishes for slabs.

In order to calculate $E_{\rm Coul}$, we used the Wigner-Seitz
approximation. In the case of spheres, bubbles, rods and tubes,
Wigner-Seitz cells were approximated by spheres and cylinders, of
radius $R_{\rm cell}$. In the case of slabs, Wigner-Seitz cells
were bounded by planes, with $R_{\rm cell}$ being defined as the
half-distance between plane boundaries of the cell. At given
average  nucleon (baryon) density, $n_{\rm b}$, 
and for an assumed shape of
nuclear structures, the energy density was minimized with respect
to  thermodynamic variables, under the condition of an average
charge neutrality.  
Spherical nuclei are energetically preferred over other 
nuclear shapes, and also over homogeneous $npe$ matter,  down to
$n_{\rm edge}=0.076~{\rm fm^{-3}}$. Within our set
of possible nuclear shapes therefore, the ground state of neutron-star crust
contains spherical nuclei only. 
A  detailed study of the bottom layers of the inner crust, 
including the determination of its bottom edge, is presented 
in Douchin \& Haensel (2000).
%
\subsection{The liquid core}
%
For $n_{\rm edge}<n_{\rm b}<2 n_0$ neutron star matter 
is expected to be a homogeneous 
plasma of neutrons, protons and electrons, and  
 - above the threshold density for
the appearance of muons (when electron chemical potential $\mu_{\rm
e}>m_\mu c^2=105.7$~MeV) - also negative muons. 
 Such a
$npe\mu$ model of dense matter is expected to be valid at not
too a high density (say, $n\la 3n_0$).  
At still higher densities one may contemplate possibility 
of the appearance of hyperons. However, in
view of a lack of detailed knowledge of the hyperon-nucleon and
hyperon-hyperon interactions, we prefer to extrapolate the
$npe\mu$ model to higher densities (this is the approach used
also in Wiringa et al. (1988), and in more recent calculation 
of Akmal et al. (1999)). 
 
The total energy density of the $npe\mu$ matter, 
$E$ (which includes rest energy of
matter constituents), is a sum of the nucleon contribution,
$E_{\rm N}(n_{\rm n},n_{\rm p})$, and of the lepton one. We have
%
\begin{eqnarray}
\label{eq:E1}
   E(n_n,n_p,n_e,n_\mu) &=& E_{\rm N}(n_n,n_p)
             + n_n m_n c^2 + n_p m_p c^2\cr
             &~& + E_e(n_e)+E_\mu(n_\mu)
\end{eqnarray}
%
where $E_e,~E_\mu$  are the energy densities,  
and  $n_e$ and $n_\mu$ are number densities of electrons and muons
respectively.
Coulomb contributions are negligible compared to the kinetic 
energies of leptons and therefore they can be treated as free
Fermi gases. 
 Mass density of matter is 
$\rho = E/c^2$.
 
The equilibrium of the $npe\mu$ matter with respect to weak interactions 
implies relations involving chemical potentials of matter constituents,
%
\begin{equation} 
\label{eq:munmu}
   \mu_n = \mu_p+\mu_e~ ,~~~~\mu_\mu = \mu_e
\end{equation}
where
%
\begin{equation} 
\label{eq:uj}
   \mu_{\rm j} = {\partial E\over
    \partial n_{\rm j}}~,~~~~{\rm j}=n,~p,~e,~\mu~. 
\end{equation}
%
At given $n_{\rm b}=n_n+n_p$, charge neutrality,
$n_ p=n_e+n_\mu$, combined with  Eq.(\ref{eq:munmu}), 
yield the equilibrium fractions 
$x_{\rm j}=n_{\rm j}/n_{\rm b}$. Using then Eq.(\ref{eq:E1}), 
one calculates the equilibrium (ground-state) value of 
 $E(n_{\rm b})$
(minimum  $E$ at given $n_{\rm b}$). This
gives a one-parameter EOS of $npe\mu$ matter,
 
%
\begin{equation}
\label{eq:rhonb}
   \rho(n_{\rm b}) = {E(n_{\rm b})\over c^2}~ ,~~~~ 
   P(n_{\rm b})    = n_{\rm b}^2{{\rm d}
     \over {\rm d}n_{\rm b}}
\left({E(n_{\rm b})\over n_{\rm b}}\right)~. 
\end{equation}
%

\section{Structure and equation of state of the crust}

\begin{table*}
\caption{\label{crust.comp} Structure and composition of the 
inner neutron-star crust (ground state) calculated within the Compressible 
Liquid Drop Model with SLy effective nucleon-nucleon interaction. 
$X_n$ is the fraction of nucleons  in the neutron gas outside nuclei. 
Upper part with $X_n=0$ corresponds to a  shell 
of the outer crust, just above  the neutron drip surface in 
the neutron-star interior, and calculated within the same model. 
 The equivalent proton and neutron radii, $R_p$ and $R_n$, are 
defined in the text. Wigner-Seitz cell radius and fraction of volume 
occupied by nuclear matter (equal to that occupied by  protons) 
are denoted by $R_{\rm cell}$ and $u$, respectively. 
 } 
\begin{tabular}{|cccccccc|} 
 \hline 
  $ n_{\rm b} $  &  $Z$ & 
$A$ & $ X_n $ & $  R_p $ & $ R_n  $& $ R_{\rm cell}  $ & $u$ \\
   $({\rm fm^{-3}})$   &   & 
 &  &   (fm)   & (fm)  & (fm)  & (\%)\\ 
 \hline 

1.2126 E-4&42.198&130.076&0.0000&   5.451& 5.915&  63.503&   0.063 \\ 
1.6241 E-4&42.698&135.750&0.0000&   5.518& 6.016&  58.440&   0.084 \\ 
1.9772 E-4&43.019&139.956&0.0000&   5.565& 6.089&  55.287&   0.102 \\ 
\cline{1-8}
2.0905 E-4&43.106&141.564&0.0000&   5.578& 6.111&  54.470&   0.107 \\ 
2.2059 E-4&43.140&142.161&0.0247&   5.585& 6.122&  54.032&   0.110 \\ 
2.3114 E-4&43.163&142.562&0.0513&   5.590& 6.128&  53.745&   0.113 \\ 
2.6426 E-4&43.215&143.530&0.1299&   5.601& 6.145&  53.020&   0.118 \\ 
3.0533 E-4&43.265&144.490&0.2107&   5.612& 6.162&  52.312&   0.123 \\ 
3.5331 E-4&43.313&145.444&0.2853&   5.623& 6.179&  51.617&   0.129 \\ 
4.0764 E-4&43.359&146.398&0.3512&   5.634& 6.195&  50.937&   0.135 \\ 
4.6800 E-4&43.404&147.351&0.4082&   5.645& 6.212&  50.269&   0.142 \\ 
5.3414 E-4&43.447&148.306&0.4573&   5.656& 6.228&  49.615&   0.148 \\ 
6.0594 E-4&43.490&149.263&0.4994&   5.667& 6.245&  48.974&   0.155 \\ 
7.6608 E-4&43.571&151.184&0.5669&   5.690& 6.278&  47.736&   0.169 \\ 
1.0471 E-3&43.685&154.094&0.6384&   5.725& 6.328&  45.972&   0.193 \\ 
1.2616 E-3&43.755&156.055&0.6727&   5.748& 6.362&  44.847&   0.211 \\ 
1.6246 E-3&43.851&159.030&0.7111&   5.784& 6.413&  43.245&   0.239 \\ 
2.0384 E-3&43.935&162.051&0.7389&   5.821& 6.465&  41.732&   0.271 \\ 
2.6726 E-3&44.030&166.150&0.7652&   5.871& 6.535&  39.835&   0.320 \\ 
3.4064 E-3&44.101&170.333&0.7836&   5.923& 6.606&  38.068&   0.377 \\ 
4.4746 E-3&44.155&175.678&0.7994&   5.989& 6.698&  36.012&   0.460 \\ 
5.7260 E-3&44.164&181.144&0.8099&   6.059& 6.792&  34.122&   0.560 \\ 
7.4963 E-3&44.108&187.838&0.8179&   6.146& 6.908&  32.030&   0.706 \\ 
 \hline 
 \end{tabular} 
\end{table*}
\begin{table*}
\caption{\label{crust.comp.cont} 
Structure and composition of the 
inner neutron-star crust - continued. 
Last line corresponds to the bottom edge of the 
inner crust.
} 
\begin{tabular}{|cccccccc|} 
 \hline 
  $ n_{\rm b} $  &  $Z$ & 
$A$ & $ X_n $ & $  R_p $ & $ R_n  $& $ R_{\rm cell}  $ & $u$ \\
   $({\rm fm^{-3}})$   &   & 
 &  &   (fm)   & (fm)  & (fm)  & (\%)\\ 
 \hline 

9.9795 E-3&43.939&195.775&0.8231&   6.253& 7.048&  29.806&   0.923 \\ 
1.2513 E-2&43.691&202.614&0.8250&   6.350& 7.171&  28.060&   1.159 \\ 
1.6547 E-2&43.198&211.641&0.8249&   6.488& 7.341&  25.932&   1.566 \\ 
2.1405 E-2&42.506&220.400&0.8222&   6.637& 7.516&  24.000&   2.115 \\ 
2.4157 E-2&42.089&224.660&0.8200&   6.718& 7.606&  23.106&   2.458 \\ 
2.7894 E-2&41.507&229.922&0.8164&   6.825& 7.721&  22.046&   2.967 \\ 
3.1941 E-2&40.876&235.253&0.8116&   6.942& 7.840&  21.053&   3.585 \\ 
3.6264 E-2&40.219&240.924&0.8055&   7.072& 7.967&  20.128&   4.337 \\ 
3.9888 E-2&39.699&245.999&0.7994&   7.187& 8.077&  19.433&   5.058 \\ 
4.4578 E-2&39.094&253.566&0.7900&   7.352& 8.231&  18.630&   6.146 \\ 
4.8425 E-2&38.686&261.185&0.7806&   7.505& 8.372&  18.038&   7.202 \\ 
5.2327 E-2&38.393&270.963&0.7693&   7.685& 8.538&  17.499&   8.470 \\ 
5.6264 E-2&38.281&283.993&0.7553&   7.900& 8.737&  17.014&  10.011 \\ 
6.0219 E-2&38.458&302.074&0.7381&   8.167& 8.987&  16.598&  11.914 \\ 
6.4183 E-2&39.116&328.489&0.7163&   8.513& 9.315&  16.271&  14.323 \\ 
6.7163 E-2&40.154&357.685&0.6958&   8.853& 9.642&  16.107&  16.606 \\ 
7.0154 E-2&42.051&401.652&0.6699&   9.312&10.088&  16.058&  19.501 \\ 
7.3174 E-2&45.719&476.253&0.6354&   9.990&10.753&  16.213&  23.393 \\ 
7.5226 E-2&50.492&566.654&0.6038&  10.701&11.456&  16.557&  26.996 \\ 
7.5959 E-2&53.162&615.840&0.5898&  11.051&11.803&  16.772&  28.603 \\ 
 \hline 
 \end{tabular} 
\end{table*}
\begin{table*}
\caption{\label{crust.EOS} Equation of state of the inner crust. 
First line corresponds to the neutron drip point, as 
calculated within the Compressible Liquid Drop Model. 
Last line corresponds to the bottom edge of the crust. 
} 
 \begin{tabular}{|cccc|cccc|} 
 \hline 
  $   n_{\rm b} $ & $ {\rm  \rho} $ & $P$  & $ \Gamma$ &
$   n_{\rm b} $ & $ {\rm  \rho} $ & $P$ & $ \Gamma$  \\ 
  (${\rm fm^{-3} }$)  & (${\rm g~cm^{-3}}$) &    
(${\rm erg~cm^{-3}}$) &
 & (${\rm fm^{-3} }$)  & (${\rm g~cm^{-3}}$) &     
            (${\rm erg~cm^{-3}}$)  & \\ 
 \hline 
2.0905 E-4&3.4951 E11&6.2150 E29&1.177&9.9795 E-3&1.6774 E13&3.0720 E31&1.342 \\ 
2.2059 E-4&3.6883 E11&6.4304 E29&0.527&1.2513 E-2&2.1042 E13&4.1574 E31&1.332 \\ 
2.3114 E-4&3.8650 E11&6.5813 E29&0.476&1.6547 E-2&2.7844 E13&6.0234 E31&1.322 \\ 
2.6426 E-4&4.4199 E11&6.9945 E29&0.447&2.1405 E-2&3.6043 E13&8.4613 E31&1.320 \\ 
3.0533 E-4&5.1080 E11&7.4685 E29&0.466&2.4157 E-2&4.0688 E13&9.9286 E31&1.325 \\ 
3.5331 E-4&5.9119 E11&8.0149 E29&0.504&2.7894 E-2&4.7001 E13&1.2023 E32&1.338 \\ 
4.0764 E-4&6.8224 E11&8.6443 E29&0.554&3.1941 E-2&5.3843 E13&1.4430 E32&1.358 \\ 
4.6800 E-4&7.8339 E11&9.3667 E29&0.610&3.6264 E-2&6.1153 E13&1.7175 E32&1.387 \\ 
5.3414 E-4&8.9426 E11&1.0191 E30&0.668&3.9888 E-2&6.7284 E13&1.9626 E32&1.416 \\ 
6.0594 E-4&1.0146 E12&1.1128 E30&0.726&4.4578 E-2&7.5224 E13&2.3024 E32&1.458 \\ 
7.6608 E-4&1.2831 E12&1.3370 E30&0.840&4.8425 E-2&8.1738 E13&2.6018 E32&1.496 \\ 
1.0471 E-3&1.7543 E12&1.7792 E30&0.987&5.2327 E-2&8.8350 E13&2.9261 E32&1.536 \\ 
1.2616 E-3&2.1141 E12&2.1547 E30&1.067&5.6264 E-2&9.5022 E13&3.2756 E32&1.576 \\ 
1.6246 E-3&2.7232 E12&2.8565 E30&1.160&6.0219 E-2&1.0173 E14&3.6505 E32&1.615 \\ 
2.0384 E-3&3.4178 E12&3.7461 E30&1.227&6.4183 E-2&1.0845 E14&4.0509 E32&1.650 \\ 
2.6726 E-3&4.4827 E12&5.2679 E30&1.286&6.7163 E-2&1.1351 E14&4.3681 E32&1.672 \\ 
3.4064 E-3&5.7153 E12&7.2304 E30&1.322&7.0154 E-2&1.1859 E14&4.6998 E32&1.686 \\ 
4.4746 E-3&7.5106 E12&1.0405 E31&1.344&7.3174 E-2&1.2372 E14&5.0462 E32&1.685 \\ 
5.7260 E-3&9.6148 E12&1.4513 E31&1.353&7.5226 E-2&1.2720 E14&5.2856 E32&1.662 \\ 
7.4963 E-3&1.2593 E13&2.0894 E31&1.351&7.5959 E-2&1.2845 E14&5.3739 E32&1.644 \\ 
 \hline 
 \end{tabular}
\end{table*}
\subsection{The outer crust}
Our calculations have been limited to  $\rho>10^6~{\rm g~cm^{-3}}$. 
The neutron star envelope with $\rho<10^6~{\rm g~cm^{-3}}$ has 
a tiny mass $\sim 10^{-10}~{\rm M}_\odot$, and moreover its 
composition and structure can be influenced by various factors, 
such as accretion of interstellar matter, and/or the presence of 
strong magnetic field.
\footnote{A more precise condition for the degenerate neutron star
 envelope to be unaffected by the magnetic field is 
$\rho>2.2~10^5 (A/Z)(B/10^{13}~{\rm G})^{3/2}~{\rm g/cm^3}$, where 
$A$ and $Z$ are mass number and atomic number of the nuclei in 
the plasma (Yakovlev \& Kaminker 1994).}

One has to be aware of the simplications and approximations 
inherent to  the CLDM. While the parameters of this model are 
determined in quantum-mechanical many-body calculation, the 
model itself is {\it par excellence} classical. 
It does not exhibit therefore the {\it shell effects} corresponding to the 
closure of proton or neutron shells in nuclei or the effect 
of neutron or proton pairing.  Shell effects 
imply particularly strong binding of nuclei with ``magic numbers''
of $Z=28$ and $N=50,~82$.  Consequently (except at lowest density) 
 nuclei present in the ground state of outer crust are expected 
to have $Z=28$ (at lower density) or $N=50,~82$ (at higher density)
(Baym et al. 1971b, Haensel \& Pichon 1994). This feature is 
absent in the CLDM EOS, which additionally treats  $Z$ and $A$ 
as continuous variables.  As a result, CLDM EOS at 
$\rho<\rho_{\rm ND}$ is {\it softer}
 and has lower value of neutron drip density 
(real nuclei are stabilized against neutron drip by the 
pairing and the shell 
 effects for neutrons)
 than that based on experimental nuclear masses. 

Actually, for $\rho\la 10^{11}~{\rm g~cm^{-3}}$, the EOS 
of the ground state of the outer crust can be reliably determined using 
{\it experimental} masses of neutron-rich nuclei (Haensel \& Pichon 1994). 
The last ``experimental'' nucleus present in the ground state of the 
outer crust is doubly magic $^{78}{\rm Ni}$ ($N=50,~Z=28$), with 
proton fraction $Z/A=0.36$. Then, up to neutron drip density, 
the EOS can be quite reliably determined using the extrapolation 
of nuclear masses,  beyond the experimentally available region, 
via semiempirical nuclear mass formulae (Haensel \& Pichon 1994). 
Within such an approach, which makes maximal use of 
experimental nuclear data, neutron drip takes place at  
$4.3~10^{11}~{\rm g~cm^{-3}}$ and for $Z/A=0.30$. 
In view of this, we suggest   
 to replace, in neutron star calculations, 
 the SLy EOS at $\rho<\rho_{\rm ND}=
4.3~10^{11}~{\rm g~cm^{-3}}$ by that of Haensel \& Pichon (1994), 
and to match it with the SLy EOS for the inner crust above 
$\rho_{\rm ND}$. This is what we did in our calculations 
of neutron star structure.  
\subsection{The inner crust}
The values of $A$, $Z$, fraction of nucleons in neutron gas outside nuclei, 
and geometrical 
parameters 
characterizing lattice of nuclei in the inner crust are displayed in 
Tables 1, 2. 
 The crust-liquid core transition takes place
not because nuclei grow in size but rather because they become
closer and closer. The filling fraction $u$ grows rapidly for
$n_{\rm b}$ approaching $n_{\rm edge}$ from only 5 \% at
$n_{\rm b}=0.04~{\rm fm^{-3}}$ to nearly 30 \% at the bottom edge.  
Still, less than 30\% of the volume is filled by nuclear matter at
the crust-liquid
 core transition point. 
Finally, let us mention that no  proton drip 
occurs in the ground state of the
crust.

As we see in Tables 1 and 2, the number of
nucleons in a nucleus, $A$, grows monotonically 
with increasing density and reaches about
600 at the edge of the crust. However, the number of protons
changes rather weakly, from $Z\simeq 40$ near  neutron drip, to
$Z\simeq 50$ near the edge of the crust. Our results for $Z$ of
spherical nuclei are  similar to those  obtained in 
Ravenhall et al. (1972) and Oyamatsu (1993), 
 but are somewhat higher than those obtained using a
relativistic mean-field model in 
Sumiyoshi et al. (1995). The problem of stability of nuclei 
in the bottom layer of the inner crust with respect to fission
was discussed in Douchin \& Haensel (2000). 

Actually,
under conditions of thermodynamic equilibrium, transition from the
crust to the uniform liquid takes place at a constant pressure,
and is accompanied by a density jump (first order phase
transition). Using Maxwell construction, we find that the edge of
the crust has density $n_{\rm edge}=0.076~{\rm fm^{-3}}$, and
coexists there with uniform $npe$ matter of the density higher by
$\Delta\rho/\rho_{\rm edge}\simeq \Delta n_{\rm b}/n_{\rm edge}
=1.4\%$. Crust-liquid core
transition is therefore a very weak first-order phase transition; it takes
place at $P_{\rm edge}=5.37~10^{32}~{\rm erg~cm^{-3}}$.
%
\section{Composition and equation of state of neutron star core}
Composition of the liquid core is given in Table 4. 
Muons appear at $n_{\rm b}=0.12~{\rm fm^{-3}}$. 
Particularly important for the rate of the neutron star 
cooling 
is the relation betwen $x_n$ and $x_p$, $x_e$, and $x_\mu$. 
Namely, the powerful direct Urca process of neutrino emission, 
$n\longrightarrow p + e + {\bar\nu}_e$, 
$p + e \longrightarrow n + \nu_e$, is allowed only if the values 
of the Fermi momenta in the $npe\mu$ matter satisfy the ``triangle
condition''
 $p_{{\rm F}n}<p_{{\rm F}p}+p_{{\rm F}e}$. 
Similarly, direct Urca process involving muons, 
$n\longrightarrow p + \mu^- + {\bar \nu}_\mu$, 
$p + \mu^- \longrightarrow n + \nu_\mu$, is allowed only if 
 $p_{{\rm F}n}<p_{{\rm F}p}+p_{{\rm F}\mu}$ (Lattimer et al. 1991).  
The ``triangle conditions'' can expressed in terms of the 
particle fractions $x_{\rm j}$ as 
\begin{eqnarray}
{\rm electron~direct~Urca:}~~~
x_n^{1\over 3}< x_p^{1\over 3} &+& x_e^{1\over 3}~,\cr\cr
{\rm muon~direct~Urca:}~~~
x_n^{1\over 3}< x_p^{1\over 3} &+& x_\mu^{1\over 3}~,
\label{Xj.Urca}
\end{eqnarray}
where we have used $p_{{\rm Fj}}=\hbar(n_{\rm j}/3\pi^2)^{1/3}$. 
The threshold density above which electron direct Urca 
process is allowed is $1.35~{\rm fm^{-3}}$, and that for the muon 
direct Urca process is somewhat higher, $1.44~{\rm fm^{-3}}$. 
 The electron direct Urca threshold for our EOS is much higher than 
$0.78~{\rm fm^{-3}}$, obtained  
for the APR model. However, in contrast to proton fraction at $n_{\rm b}\la 
n_0$, our values of $x_p$ at higher density should be taken with a
grain of salt: this feature of our model is based on an extrapolation 
from  nuclear densities, and - in contrast to the EOS itself - it was not 
subjected to a reliable constraint.

\begin{table*}
\caption{\label{compos.core} Composition  of the liquid core. 
Fractions of particles are defined as 
$x_{\rm j}=n_{\rm j}/n_{\rm b}$. Neutron fraction can be calculated 
using $x_n=1-x_p$.}
\begin{tabular}{|cccc|cccc|} 
  \hline 
$ n_{\rm b} $ & $ x_p $ &$  x_e $ & $ x_{\mu} $ & 
$ n_{\rm b} $ & $ x_p $ &$  x_e$ & $ x_{\mu}  $   \\ 
$ ({\rm fm^{-3}}) $ & (\%) & (\%) & 
(\%) & $({\rm fm^{-3}})$ & (\%)  & (\%)  &  (\%) \\
 \hline 
0.0771& 3.516& 3.516& 0.000& 0.490& 7.516& 4.960& 2.556 \\ 
0.0800& 3.592& 3.592& 0.000& 0.520& 7.587& 4.954& 2.634 \\ 
0.0850& 3.717& 3.717& 0.000& 0.550& 7.660& 4.952& 2.708 \\ 
0.0900& 3.833& 3.833& 0.000& 0.580& 7.736& 4.955& 2.781 \\ 
0.1000& 4.046& 4.046& 0.000& 0.610& 7.818& 4.964& 2.854 \\ 
0.1100& 4.233& 4.233& 0.000& 0.640& 7.907& 4.979& 2.927 \\ 
0.1200& 4.403& 4.398& 0.005& 0.670& 8.003& 5.001& 3.002 \\ 
0.1300& 4.622& 4.521& 0.101& 0.700& 8.109& 5.030& 3.079 \\ 
0.1600& 5.270& 4.760& 0.510& 0.750& 8.309& 5.094& 3.215 \\ 
0.1900& 5.791& 4.896& 0.895& 0.800& 8.539& 5.178& 3.361 \\ 
0.2200& 6.192& 4.973& 1.219& 0.850& 8.803& 5.284& 3.519 \\ 
0.2500& 6.499& 5.014& 1.485& 0.900& 9.102& 5.410& 3.692 \\ 
0.2800& 6.736& 5.031& 1.705& 0.950& 9.437& 5.557& 3.880 \\ 
0.3100& 6.920& 5.034& 1.887& 1.000& 9.808& 5.726& 4.083 \\ 
0.3400& 7.066& 5.026& 2.040& 1.100&10.663& 6.124& 4.539 \\ 
0.3700& 7.185& 5.014& 2.170& 1.200&11.661& 6.602& 5.060 \\ 
0.4000& 7.283& 4.999& 2.283& 1.300&12.794& 7.151& 5.643 \\ 
0.4300& 7.368& 4.984& 2.383& 1.400&14.043& 7.762& 6.281 \\ 
0.4600& 7.444& 4.971& 2.473& 1.500&15.389& 8.424& 6.965 \\ 
 \hline 
 \end{tabular}
\end{table*}
 
\begin{table*}
\caption{\label{eos.core} 
Equation of state of the liquid neutron-star core}
 \begin{tabular}{|cccc|cccc|} 
 \hline 
  $   n_{\rm b} $ & $\rho$ & $P$  & $ \Gamma$ &
 $   n_{\rm b} $ & $\rho$ & $P$ & $ \Gamma $  \\ 
  (${\rm fm^{-3} }$)  & (${\rm g~cm^{-3}}$) &    
(${\rm erg~cm^{-3}}$) &
 & (${\rm fm^{-3} }$)  & (${\rm g~cm^{-3}}$) &     
            (${\rm erg~cm^{-3}}$)  & \\ 
 \hline 
  0.0771&1.3038 E14&5.3739 E32&2.159&  0.4900&8.8509 E14&1.0315 E35&2.953 \\ 
  0.0800&1.3531 E14&5.8260 E32&2.217&  0.5200&9.4695 E14&1.2289 E35&2.943 \\ 
  0.0850&1.4381 E14&6.6828 E32&2.309&  0.5500&1.0102 E15&1.4491 E35&2.933 \\ 
  0.0900&1.5232 E14&7.6443 E32&2.394&  0.5800&1.0748 E15&1.6930 E35&2.924 \\ 
  0.1000&1.6935 E14&9.9146 E32&2.539&  0.6100&1.1408 E15&1.9616 E35&2.916 \\ 
  0.1100&1.8641 E14&1.2701 E33&2.655&  0.6400&1.2085 E15&2.2559 E35&2.908 \\ 
  0.1200&2.0350 E14&1.6063 E33&2.708&  0.6700&1.2777 E15&2.5769 E35&2.900 \\ 
  0.1300&2.2063 E14&1.9971 E33&2.746&  0.7000&1.3486 E15&2.9255 E35&2.893 \\ 
  0.1600&2.7223 E14&3.5927 E33&2.905&  0.7500&1.4706 E15&3.5702 E35&2.881 \\ 
  0.1900&3.2424 E14&5.9667 E33&2.990&  0.8000&1.5977 E15&4.2981 E35&2.869 \\ 
  0.2200&3.7675 E14&9.2766 E33&3.025&  0.8500&1.7302 E15&5.1129 E35&2.858 \\ 
  0.2500&4.2983 E14&1.3668 E34&3.035&  0.9000&1.8683 E15&6.0183 E35&2.847 \\ 
  0.2800&4.8358 E14&1.9277 E34&3.032&  0.9500&2.0123 E15&7.0176 E35&2.836 \\ 
  0.3100&5.3808 E14&2.6235 E34&3.023&  1.0000&2.1624 E15&8.1139 E35&2.824 \\ 
  0.3400&5.9340 E14&3.4670 E34&3.012&  1.1000&2.4820 E15&1.0609 E36&2.801 \\ 
  0.3700&6.4963 E14&4.4702 E34&2.999&  1.2000&2.8289 E15&1.3524 E36&2.778 \\ 
  0.4000&7.0684 E14&5.6451 E34&2.987&  1.3000&3.2048 E15&1.6876 E36&2.754 \\ 
  0.4300&7.6510 E14&7.0033 E34&2.975&  1.4000&3.6113 E15&2.0679 E36&2.731 \\ 
  0.4600&8.2450 E14&8.5561 E34&2.964&  1.5000&4.0498 E15&2.4947 E36&2.708 \\ 
 \hline 
\end{tabular}
\end{table*}
Equation of state of the liquid core is  given in Table 5. 
Its properties will be discussed in Sect.5. Here we will 
restrict ourselves to a comment referring to its 
practical use in neutron star calculations. Tiny 
density jump between core and crust is not relevant for 
the applications to calculations of the neutron star 
structure (albeit it can play some role in neutron star 
dynamics). One can remove the first line of Table 5 and then 
match the resulting EOS of the core to that of the inner 
crust, given in Table 3. However, one can also  
remove last line if Table 3, and then match the EOS 
of the inner crust to that of the inner core, given in 
Table 5. In practice,  the difference in neutron star structure,  
resulting from the difference in these two prescriptions, 
is negligibly small.  
%
\section{Properties of the EOS}
Using the CLDM of dense matter, based on the SLy effective 
nucleon-nucleon interaction, we determined the EOS of 
cold neutron-star matter in the interval of density 
from $10^8~{\rm g~cm^{-3}}$ to 
 $4~10^{15}~{\rm g~cm^{-3}}$. This EOS is displayed in 
Fig. 1, where for the sake of clarity of presentation 
at higher density, the lower-density limit has been set 
at $10^{10}~{\rm g~cm^{-3}}$. Our EOS covers three main 
regions of neutron-star interior: outer crust, inner crust and 
liquid core. In the displayed region of the outer crust the 
EOS is well approximated by a polytrope, with nearly constant adiabatic 
index (see below). Just after neutron drip, the EOS softens considerably, 
gradually  stiffens in the higher-density part of the inner crust 
and than stiffens considerably after crossing 
the crust-core interface.  
\subsection{Adiabatic index $\Gamma$}
An important dimensionless parameter characterizing the stiffness  
of the EOS at given density is the adiabatic index, defined by
%
\begin{equation}
\Gamma=
{n_{\rm b}\over P} {{\rm d}P\over {\rm d}n_{\rm b}}= 
{\rho + {P/c^2}\over P} {{\rm d}P\over {\rm d}\rho} ~.
\label{Gamma}
\end{equation}
%
The three main regions of the neutron star interior are 
characterized by distinct behavior of $\Gamma$, 
displayed in Fig. 2. Precise values of $\Gamma$ are 
given in Tables \ref{crust.EOS} and \ref{eos.core}. 
In the outer crust, the value 
of $\Gamma$ depends quite weakly on density. 
In should be mentioned that a weak, smooth decrease of 
$\Gamma$ in the  higher-density part of the outer 
crust, displayed in Fig. 2, is an artifact of the CLDM of dense 
matter. Had we used a model of the outer crust, based on the 
$A,Z$ table of nuclear masses (experimental where available, 
calculated using semi-empirical mass formula elsewhere), 
we would get a sequence of shells with fixed $A,Z$, and 
density jump at the neighboring shells interface (Baym et al. 
1971b, for a recent calculation see Haensel \& Pichon 1994). 
Above $10^8~{\rm g~cm^{-3}}$, adiabatic index within 
each $A,Z$ shell would be    $\Gamma\simeq 4/3$. 
This can be easily understood, because the  main component 
of the matter pressure comes from 
the ultrarelativistic electron gas, 
and the most important correction from Coulomb (lattice) term 
in the energy  density, both of which  
behave as $\propto (Z\rho/A)^{4/3}$ 
(Baym et al. 1971b). Notice that for the outer crust $\rho$ 
is to a very good approximation, proportional to $n_{\rm b}$. 
In the CLDM, with continuous variables $A,Z$, density 
jumps do not appear, and decrease of $\Gamma$ 
below $4/3$ results mostly from the monotonic, smooth 
increase of $A$ with increasing density. Actually, such 
a behavior within the CLDM simulates ``averaging'' 
of the  value of $\Gamma$ over the density 
jumps, which effectively softens (as first-order phase transitions 
between shells with  different nuclides should do) the EOS of the 
outer crust. 

Neutron drip at $\rho_{\rm ND}$ implies a dramatic drop 
in $\Gamma$, which corresponds to strong softening 
of the EOS. Density stays continuous at the neutron drip point, 
with low-density dripped neutrons contributing to $n_{\rm b}$ and 
$\rho$, but exerting a very small pressure, and moreover being  in 
phase equilibrium with nuclear matter of nuclei. 
Consequently, $\Gamma$ drops  by more than a factor of 
two, a sizable part of this drop  occurring via discontinuous drop at 
$\rho_{\rm ND}$, characteristic of a second-order phase transition. 
After this initial dramatic drop, matter stiffens, because 
pressure of neutron gas inscreases. The actual 
value of $\Gamma$ results from an 
interplay of several factors, with stiffening  due 
to Fermi motion and neutron-neutron repulsion in 
dripped non-relativistic 
neutron gas and,   countering this,  softening 
Coulomb (lattice) contribution,  a rather soft contribution 
of ultrarelativistic electron gas, and a softening effect 
of neutron gas - nuclear matter coexistence. 
As one sees in Fig. 2, $\Gamma$ reaches 
the value of  
 about 1.6 near the bottom edge of the inner crust, only slightly lower 
than 5/3 characteristic of a non-relativistic free Fermi gas. 

At the crust-core interface, matter strongly stiffens, and 
$\Gamma$ increases 
 discontinuously,  by 0.5, to about 
2.2. This jump results from  the disappearence  of nuclei: 
a two-phase nucleon system changes into a  
single-phase one, and repulsive nucleon-nucleon 
 interaction is no longer 
countered by softening effects  resulting from the presence 
of nuclear structure  and neutron gas - nuclear matter 
phase coexistence.  With increasing density, $\Gamma$ 
grows  above 3 at $2\rho_0$, due to increasing contribution 
of repulsive nucleon interactions. 

A tiny notch appears at the muon threshold, at which $\Gamma$ 
undergoes small, but clearly visible, discontinuous drop. 
It is due to the appearance of  new fermions - muons, which  
 replace high-energy electrons (electron 
Fermi energy $\mu_e\ge 105.7$~MeV). Replacing rapidly moving electrons 
by slowly moving muons leads to a drop in  
the sound velocity (and $\Gamma$) just after the 
threshold. Because lepton contribution to pressure is 
at this density very small, the overall effect is small. 
Discontinuous 
drop in $\Gamma$ at muon threshold is characteristic 
of a second-order phase transition at which density is continuous 
but compressibility is not.  

At higher densities,  $\rho>2\rho_0$,  $\Gamma$ decreases 
slowly, which  results from the interplay of the density dependence 
of nuclear interactions and of increasing proton fraction. 
\subsection{Deviations from beta-equilibrium and $\Gamma_{\rm fr}$}
The calculation of the EOS has been done under assumption of 
full thermodynamic equilibrium. Therefore,  $\Gamma$ 
determines the response of neutron star matter to a local change 
of density when this assumption is valid. In the case 
of middle aged or old neutron stars however, the timescale of 
beta processes, which assure beta equilibrium expressed in 
Eq. \ref{eq:munmu}, 
 is many orders of magnitude 
 longer than the characteristic 
timescales of dynamical phenomena, such as stellar pulsations 
or sound waves, excited in neutron star 
interior. In such a case, proton, electron 
and muon  fractions  
in a perturbed element of neutron 
star matter cannot adjust to the instantaneous 
value of $n_{\rm b}$ because beta processes are too slow, 
and in practice the values of $x_p$, $x_e$, and $x_\mu$ 
can be considered as fixed at their unperturbed values. 
In view of this, the response of pressure to perturbation 
of density is determined by $\Gamma_{\rm fr}$, calculated 
under the condition of constant composition (Gourgoulhon 
et al. 1995). Both adiabatic indices are related by 
\begin{equation}
\Gamma=\Gamma_{\rm fr} 
+ {n_{\rm b}\over P}\sum_{\rm j} 
\left(
{\partial P\over \partial x_{\rm j}}
\right)_{n_{\rm b}}
\left(
{\partial x_{\rm j}\over \partial n_{\rm b}}
\right)_{\rm eq}~,
\label{Gammas}
\end{equation}
where the index ``eq'' indicates that the derivative has to 
be calculated assuming beta equilibrium (i.e., from the EOS), 
and ``fr'' indicates a constant (frozen) composition. 

Both $\Gamma$ and $\Gamma_{\rm fr}$ in the liquid 
interior are shown in Fig.\ref{Gamma.core}. Freezing the composition 
stiffens neutron star matter, $\Gamma_{\rm fr}>\Gamma$, 
the effect being of the order of a few percent. Another effect  
of the composition freezing is removing   of a softening 
just after the appearence of muons, because of the slowness 
of processes in which they are produced or absorbed.  
\subsection{Velocity of sound and causality}
The adiabatic sound speed is given by 
\begin{equation}
v_{\rm s}^2= \left(
{\partial P\over \partial \rho}
\right)_S~,
\label{vsound.def}
\end{equation}
where $S$ is the entropy per baryon. The value of $v_{\rm s}$ may 
become comparable to $c$ in dense neutron star core.  

Characteristic period of sound waves excited in the liquid core can 
be estimated as $\tau_{\rm s}\sim R/v_{\rm s}\sim 0.1$~ms.  
 Therefore, $\tau_{\rm s}$ is much shorter than 
the timescale of beta processes, so that in this case 
\begin{equation}
v_{\rm s}=c
\sqrt{ 
{\Gamma_{\rm fr}\over \rho c^2/P  + 1}
}~.
\label{vsound.Gamma}
\end{equation}
A {\it necessary} condition for causality in a liquid medium is  
$v_{\rm s}\le c$: sound has to be {\it subluminal}. As our 
many-body model for the nucleon component of dense matter 
is non-relativistic, it is not obvious {\it a priori} that 
the EOS we have calculated respects this condition at 
high density.  
The calculation shows, that $v_{\rm s}\le c$ is 
valid for our EOS at   
all densities relevant for neutron stars (i.e., for $\rho<
\rho_{\rm max}$, see Sect.6.1). It should be mentioned, however, 
that $v_{\rm s}\le c$ is not {\it sufficient} for causality to 
be respected by a neutron star matter model. Strict conditions 
for causality can be only derived using the kinetic theory, 
which describes all modes which can propagate in neutron star 
matter (Olson 2000). In a simplified case of a 
schematic (unrealistic) model of neutron star 
matter  a set of  conditions resulting from kinetic equations 
was obtained by Olson (2000). However, the problem of a complete 
set of causality conditions was not studied for realistic models 
of neutron star matter, and we will not attempt to solve this 
problem in the present paper.    

\begin{figure}
\resizebox{\hsize}{!}{\includegraphics{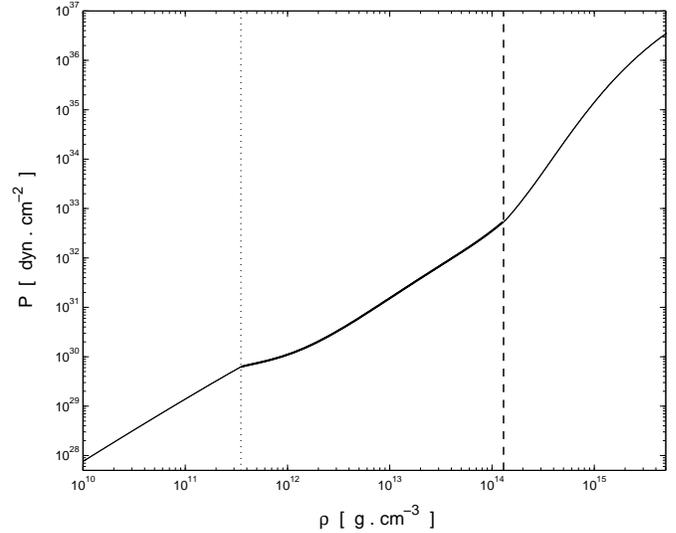}}
\caption{
The SLy EOS of the ground-state neutron star matter. 
Dotted vertical line corresponds  to the neutron drip 
and the dashed one to the crust-liquid core interface. 
  }
 \label{eos.all}
\end{figure}
\begin{figure}
\resizebox{\hsize}{!}{\includegraphics{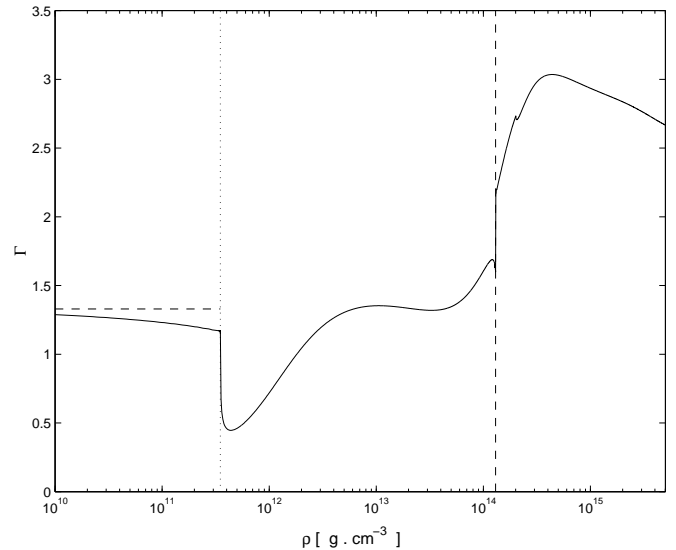}}
\caption{
Adiabatic index of the Sly EOS versus matter density. 
Matter is assumed to be in full thermodynamic equilibrium. 
Dotted vertical line: neutron drip. Dashed vertical line: 
crust - liquid core interface. Dashed horizontal line 
in the region of the outer crust is $\Gamma=1.33$ obtained 
using empirical and semi-empirical (i.e., from mass formulae) 
 masses of nuclei and 
removing the points corresponding to the 
 density jumps between shells with different 
nuclei (Baym et al. 1971b, Haensel \& Pichon 1994).   
  }
 \label{Gamma.all}
\end{figure}
\begin{figure}
\resizebox{\hsize}{!}{\includegraphics{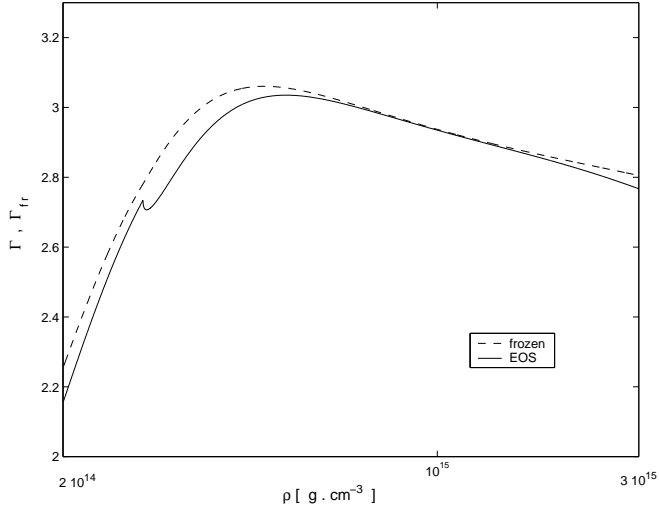}}
\caption{
Adiabatic index of the  SLy EOS of the liquid core. 
Solid line: beta equilibrium of $npe\mu$ matter 
(equilibrium composition during compression or decompression). 
Dashed line: vanishing rate of weak processes in 
$npe\mu$ matter (frozen composition during compression or 
decompression). 
  }
 \label{Gamma.core}
\end{figure}
\section{Neutron star structure. Static equilibrium 
configurations.}
Configurations of hydrostatic equilibrium of 
non-rotating neutron stars have been calculated by 
solving the Tolman-Oppenheimer-Volkoff (TOV) 
equations
%
\begin{eqnarray}
{{\rm d}P\over {\rm d}r}&=& 
- {G\rho m\over r^2} \left(1 + {P\over \rho c^2}\right)
\left(1  + {4\pi P r^3\over m c^2}\right)
\left(1  -{2Gm\over r c^2}\right)^{-1}~,\cr
{{\rm d}m\over {\rm d}r} &=& 4\pi r^2 \rho~,
\label{TOV.eq}
\end{eqnarray}
%
where $r$ is the radial coordinate in  the Schwarzschild metric. 
The TOV equations are supplemented with an equation determining 
number of baryons, $a$, within the sphere of radius $r$, 
%
\begin{equation}
{{\rm d}a\over {\rm d}r} = 
4\pi r^2 n_b \left( 1 - {2Gm\over r c^2}\right)^{-{1\over 2}}~.
\label{a.eq}
\end{equation}
%
Eqs.(\ref{TOV.eq},\ref{a.eq}) were integrated from the center of 
the configuration,  with boundary condition at $r=0$: 
$P(0)=P_{\rm c}$, $m(0)=0$, $a(0)=0$. The stellar surface 
at $r=R$ was then determined from $P(R)=0$. The total gravitational 
mass $M=m(R)$, and the total number of baryons $A=a(R)$. 

Models of cold, static neutron stars form a one-parameter 
family. They can be labeled by their central pressure, 
$P_{\rm c}$, or equivalently by their  central 
 density, $\rho_{\rm c}$.  
\begin{figure}
\resizebox{\hsize}{!}{\includegraphics{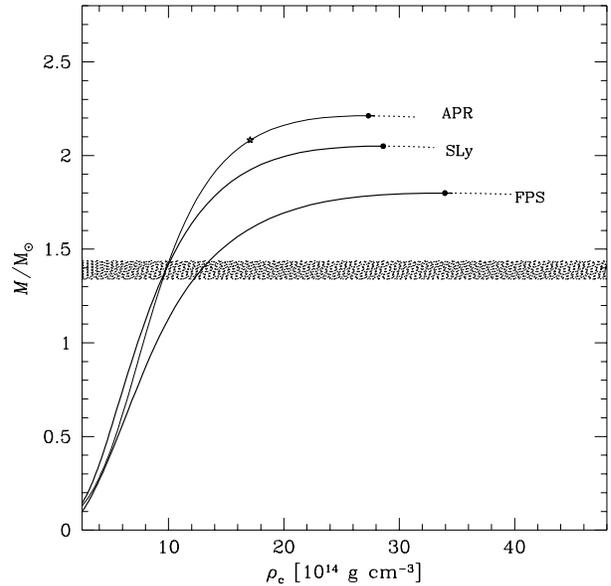}}
\caption{
Gravitational mass $M$ versus central density $\rho_{\rm c}$, 
for the SLy, FPS, and APR  EOS of dense matter. 
Maximum on the mass-central density curves is 
indicated by a filled 
circle. On the APR curve, configurations to the right of the asterisk 
contain a central core  with $v_{\rm sound}>c$.  
 Configurations to the right of the maxima  are unstable 
with respect to small radial perturbations, and are denoted 
by a dotted line. The shaded band corresponds to the range of 
precisely measured masses of 
binary radio pulsars.
  }
 \label{Mrhoc}
\end{figure}
\begin{figure}
\resizebox{\hsize}{!}{\includegraphics{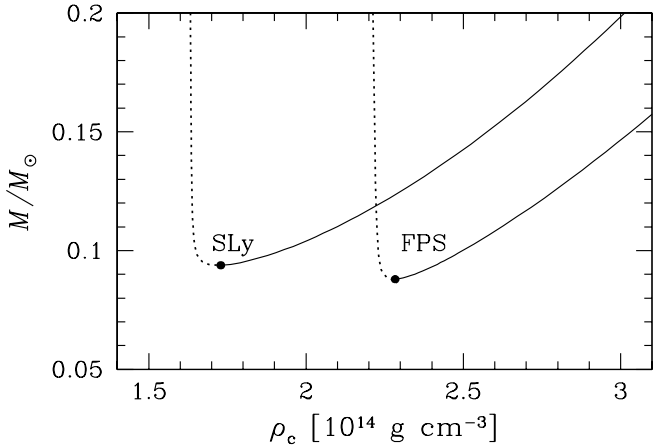}}
\caption{
Gravitational mass  versus central density, in the vicinity of 
the minimum mass,   for static neutron stars. Dotted lines - 
configurations unstable with respect to small radial 
perturbations. Minimum mass configuration is indicated 
by a filled circle. 
}
 \label{Mrhoc_min}
\end{figure}
\begin{figure}
\resizebox{\hsize}{!}{\includegraphics{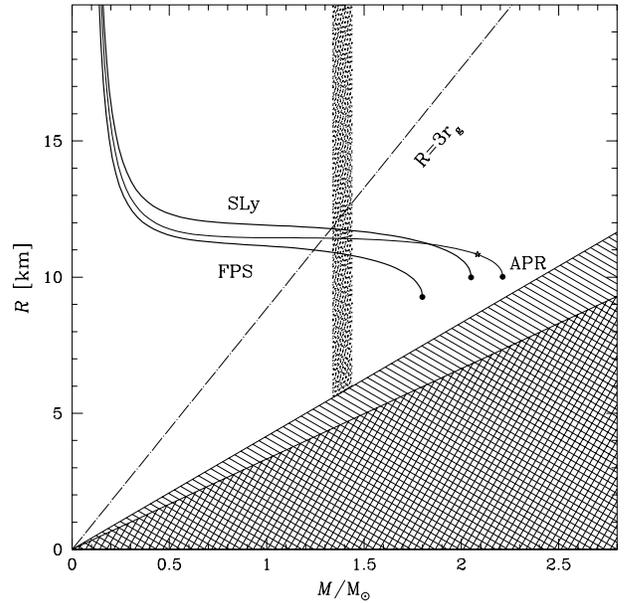}}
\caption{
Neutron star radius $R$  versus gravitational mass $M$, 
with notation as in Fig.\ref{Mrhoc}. Doubly 
hatched area is prohibited by general relativity, 
because it corresponds to $R<{9\over 8}r_{\rm g}=9GM/4c^2$ 
(for a general proof, see Weinberg 1972). All 
hatched triangle (double and single hatched) is 
prohibited by the general relativity and 
condition $v_{\rm sound}<c$ (necessary but not 
sufficient for respecting causality (Olsson 2000)) 
combined. The shaded band corresponds to the range of 
precisely measured masses of binary radio pulsars. 
  }
 \label{RM}
\end{figure}
\begin{table*}
\caption{\label{Mmax.tab}
Configuration of maximum allowable mass  for static neutron stars}
\begin{tabular}{|cccccccccc|}
\hline
 EOS & $M$ &  $R$  & $n_{\rm c}$ & $\rho_{\rm c}$ &
 $P_{\rm c}$ &  $A$  & $z_{\rm surf}$ & $E_{\rm bind}$ & $I$\\
 & $[{\rm M}_\odot]$&[km]& $[{\rm fm^{-3}}]$&
$[10^{14}~{\rm g/cm^3}]$& 
  $[{\rm 10^{36}~dyn/cm^2}]$& $[10^{57}]$& &$[10^{53}~{\rm erg}]$&
$[10^{45}~{\rm g~cm^2}]$\\ 
\hline
&&&&&&&&&\\
 SLy & 2.05 & 9.99 & 1.21 & 2.86 & 1.38 & 2.91 & 0.594 & 6.79 & 1.91\\
&&&&&&&&&\\
FPS  & 1.80  & 9.27 & 1.46  & 3.40 & 1.37 & 2.52 & 0.531 & 5.37 & 1.36\\
&&&&&&&&&\\
\hline
\end{tabular}
\end{table*}
\begin{figure}
\resizebox{\hsize}{!}{\includegraphics{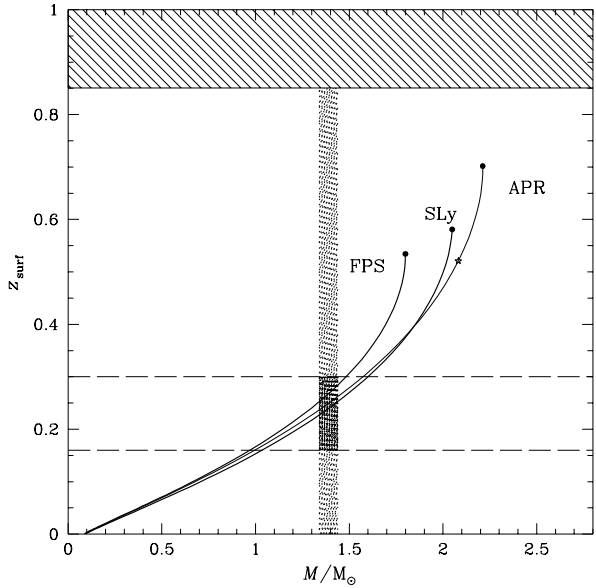}}
\caption{
Surface redshift $z_{\rm surf}$ versus gravitational mass 
$M$. Hatched area is prohibited for EOSs with $v_{\rm sound}
<c$. Shaded 
vertical band corresponds to the range of 
precisely measured masses 
of binary radio pulsars. The band limited by two dashed horizontal 
lines corresponds to the estimate of $z_{\rm surf}$ from the 
measured  spectrum of the gamma-ray burst GB 790305b.  
  }
 \label{zM}
\end{figure}
\begin{figure}
\resizebox{\hsize}{!}{\includegraphics{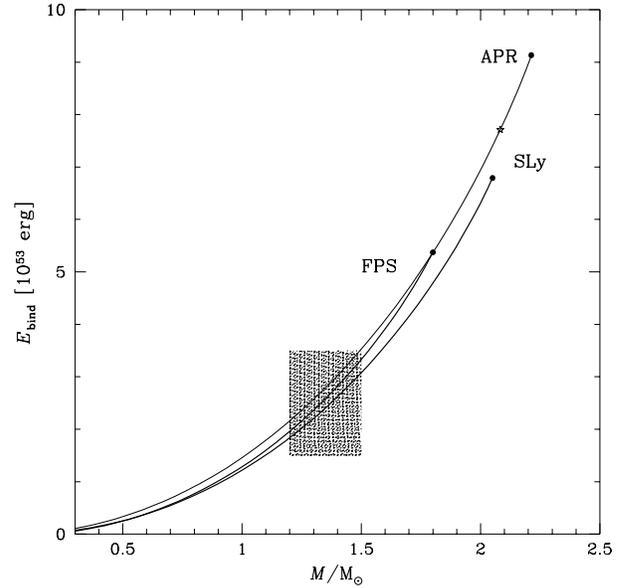}}
\caption{
 Binding energy relative to dispersed $^{56}{\rm Fe}$ versus 
gravitational mass. The shaded rectangle corresponds to the estimates 
of the total energy of the neutrino burst in SN 1987A, and 
to the estimates of the mass of neutron star formed in this 
event. 
  }
 \label{EbM}
\end{figure}
\begin{figure}
\resizebox{\hsize}{!}{\includegraphics{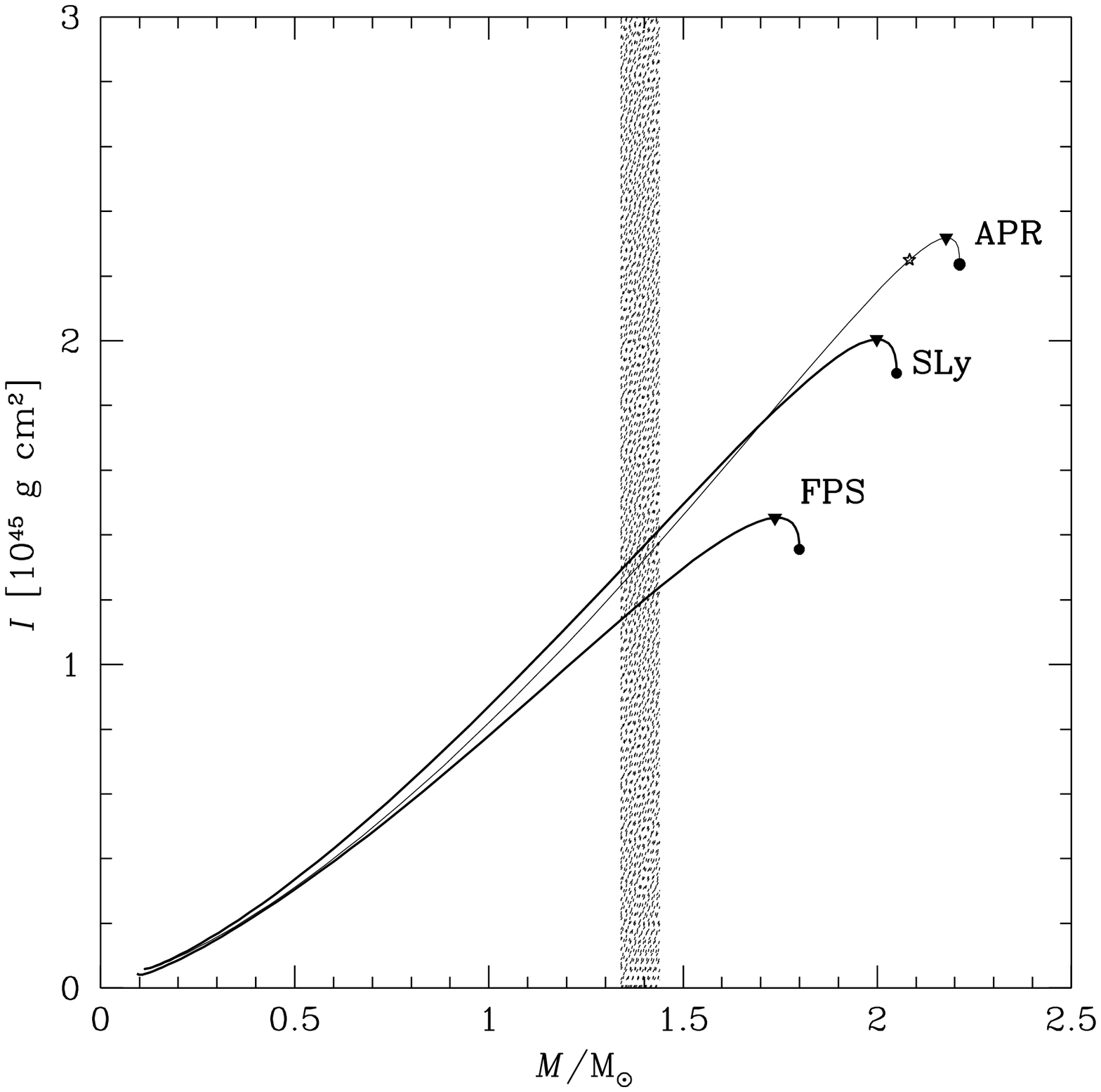}}
\caption{
Moment of inertia for slow, rigid rotation versus 
gravitational mass. The configuration with maximum $I$ is indicated 
by a filled triangle, and that of maximum mass - by a filled 
circle. A shaded band corresponds to the range of precisely 
measured masses of binary radio pulsars. 
  }
 \label{IM}
\end{figure}
\subsection{Mass, central density, and radius of neutron stars}
In Fig. 4  we show  dependence of gravitational mass on the central 
density, $\rho_{\rm c}$, for $\rho_{\rm c}>3~10^{14}~{\rm g~cm^{-3}}$, 
and compare it with that obtained for the FPS EOS. 
Actually, on the lower-density side the curve $M(\rho_{\rm c})$ 
exhibits a minimum at $M_{\rm min}\simeq 0.09~{\rm M}_\odot$, 
not shown  in the figure. The value of $M_{\rm min}$ depends 
rather weakly on the EOS. On the higher-density side, 
$M(\rho_{\rm c})$ has a maximum. The existence of $M_{\rm max}$ 
(for any EOS) is an important  consequence of general relativity. 
Configuration with $M>M_{\rm max}$ cannot exist in hydrostatic 
equilibrium and collapse into black holes. We get $M_{\rm max}=
2.05~{\rm M}_\odot$, to be compared with  $1.80~{\rm M}_\odot$ 
for a softer FPS EOS.  

Central density of the maximum allowable mass configuration 
is the maximum one which can be reached within static neutron stars. 
Models with $\rho_{\rm c}>\rho_{\rm c}(M_{\rm max})
\equiv \rho_{\rm max}$ have ${\rm d}M/{\rm d}\rho_{\rm c}<0$. 
They are therefore unstable 
 with respect to small radial perturbations 
and collapse into black holes  (see, e.g., Shapiro 
\& Teukolsky 1983). 
The maximum central density for static stable neutron stars  
is, for our EOS,  $2.9~10^{15}~{\rm g~cm^{-3}}$, to be compared with 
$3.4~10^{15}~{\rm g~cm^{-3}}$ for the FPS EOS. 
Corresponding maximum value of baryon density  
is $n_{\rm max}=1.21~{\rm fm^{-3}}\simeq 7.6n_0$, to be 
compared with $1.46~{\rm fm^{-3}}\simeq 
9.1n_0$ obtained  for the FPS EOS. Complete set of 
parameters of configuration with maximum allowable mass 
for our EOS is  presented  in Table \ref{Mmax.tab}, where 
the corresponding parameters obtained for the FPS EOS are also 
given for comparison.

 Comparison with the APR EOS is also of interest, and therefore we 
 show  the $M(\rho_{\rm c})$ curve for this EOS. 
The curve obtained for our EOS is quite close to the APR one, especially for 
$1\la M/{\rm M_\odot}\la 2$.  It should be mentioned, that 
for $\rho_{\rm c}>1.73~10^{15}~{\rm g~cm^{-3}}$ the APR neutron star 
models contain a central core with $v_{\rm sound}>c$, and should therefore 
be taken with a grain of salt. Such a problem does not arise for our EOS, 
for which $v_{\rm sound}<c$ within all stable neutron star models.

Precisely measured masses of radio pulsars in binaries with 
another neutron star span the range $1.34-1.44~{\rm M}_\odot$ 
(Thorsett \& Chakrabarty 1999), 
visualized in Fig. 4 by a shaded band. For neutron stars of such masses, 
central density is about  $1~10^{15}~{\rm g~cm^{-3}}$, slightly 
below  $4\rho_0$; this result is nearly the same as for the 
APR EOS. For the FPS EOS, 
neutron star of such a mass has 
higher central density,   about $1.3~10^{15}~{\rm g~cm^{-3}}\simeq 
5\rho_0$. 
%
\subsection{Minimum mass of neutron stars}
\begin{table*}
\caption{\label{Mmin.tab}
Configurations of minimum mass for static neutron stars}
\begin{tabular}{|cccccc|}
\hline
 EOS & $M_{\rm min}$ & $\rho_{\rm c}$ & $R$ &
 $M_{\rm core}/M$ & $R_{\rm core}$\cr
  & $[{\rm M}_\odot]$ & $[10^{14}~{\rm g/cm^3}]$ & [km] &
 &    [km] \cr
\hline
&&&&&\\
SLy  & 0.094  & 1.6  &  270  & 0.02 & 3.8\\
&&&&&\\
FPS  & 0.088  & 2.2  & 220 & 0.03 & 4.2\\
&&&&&\\
\hline
\end{tabular}
\end{table*}
With decreasing value of the central density, the mass of equilibrium 
configuration decreases. Finally, one reaches the minimum, 
$M_{\rm min}$, on the $M-\rho_{\rm c}$ curve. 
The curve $M(\rho_{\rm c})$ in the neighbourhood of $M_{\rm min}$ 
is shown in Fig.\ref{Mrhoc_min}. Parameters of the minimum mass 
configuration for static neutron stars  are given in 
Table \ref{Mmin.tab}, 
where for the sake of comparison we show also  corresponding 
values calculated for the FPS EOS. 

The value of $M_{\rm min}$ for the SLy EOS and the FPS EOS are quite 
similar: in both cases $M_{\rm min}\simeq  0.09~{\rm M}_\odot$. 
Since the SLy EOS is stiffer than the FPS one in the vicinity of 
the crust-core interface, its  $M_{\rm min}$ configuration 
is less dense and has larger radius. In both cases it has  
a small central liquid core, containing  $2\%$ of mass in the 
case of the SLy EOS and $3\%$ of  star mass in the case of a
softer FPS EOS.  

\subsection{Radius versus gravitational mass}
The radius-mass relation, obtained for our EOS 
 for static, cold neutron stars,  is shown 
in Fig. 5, where for the sake of comparison we show also 
the $R(M)$ curve for the FPS EOS. For masses between $1~{\rm M}_\odot$ 
and $M_{\rm max}=2.05~{\rm M}_\odot$, the 
neutron star radius decreases rather weakly 
with increasing mass, from 12 km to 10 km. For neutron star masses, 
measured for some binary radio pulsars (shaded band), the radius is slightly 
below 12 km. {\rm  The insensitivity of $R$ to $M$ for $1\la M/{\rm M}_\odot 
\la M_{\rm max}$ is typical of the realistic EOS without a strong softening 
at high density. The reasons for such a weak dependence of $R$ on $M$ 
have been explained by Lattimer \& Prakash (2001).}

At the same value of $M$ between  $0.5~{\rm M}_\odot$ 
and $1.5~{\rm M}_\odot$, the radius of the SLy 
neutron star is some $\sim 1~$km larger than that obtained for 
the FPS EOS; the difference increases with increasing $M$, 
and reaches 2 km at  $1.80~{\rm M}_\odot$. 
 It is of interest to compare $R(M)$ curve for our EOS 
also with that obtained for the APR EOS. For $1\la M/{\rm M}_\odot\la 2$ 
both curves are quite  similar. Notice, that highest-mass segment (to 
the right of the asterisk) of the APR curve should be treated 
with caution, because  stellar models contain there a central 
core with $v_{\rm sound}>c$. It is due to this unphysical feature 
that the APR $R(M)$ curve approaches so closely the prohibited 
hatched region of the $R-M$ plane.

For a static neutron star, general relativity predicts  that 
the circular Keplerian orbits (for test particles) 
with $r>3 r_{\rm g}$ are stable, and those 
with $r<3 r_{\rm g}$ are unstable,  
where the gravitational radius $r_{\rm g}\equiv 2GM/c^2=
2.95~M/{\rm M}_\odot$~km (see, e.g., Shapiro \& Teukolsky 
1983). 

The radius of the 
{\it marginally stable} orbit, which separates these two classes 
of orbits, is therefore $r_{\rm ms}=3r_{\rm g}=
12.4~(M/1.4{\rm M}_\odot)~$km. As we see in Fig.\ref{RM}, 
for $M\ga 1.4~{\rm M}_\odot$ we have $r_{\rm ms}>R$, and 
therefore for such neutron stars the innermost stable circular orbit 
(ISCO) is  separated from the stellar surface by 
a gap. A similar situation holds also for the FPS EOS. Let us notice, 
that the existence of a gap between the ISCO and neutron star surface
might be  important for the interpretation  of the spectra of  
the kiloherhz Quasi Periodic Oscillations observed in the X-ray 
radiation of some Low Mass X-ray Binaries (van der Klis 2000).  
\subsection{Surface redshift}
Surface redshift of photons emitted from neutron star photosphere 
is given by
%
\begin{equation}
z_{\rm surf}=\left(1 - {r_{\rm g}\over R}\right)^{-{1\over 2}}
 - 1~. 
\label{zsurf.Eq}
\end{equation}
%
 Surface redshift versus gravitational mass is 
plotted in Fig. \ref{zM}. 
At given $M$, the SLy value of $z_{\rm surf}$ is 
systematically lower than for softer FPS EOS. However, the maximum 
surface redshift for the SLy EOS, 0.59, is some 10\% higher than 
for the FPS EOS, 0.53. The larger value of $M_{\rm max}$ for 
stiffer SLy EOS plays a decisive role in determining the spacetime 
curvature close to neutron star with maximum allowable mass. 
In the range of measured values of masses of binary pulsars 
we get $z_{\rm surf}\simeq 0.22\div 0.26$, slightly higher than 
for the FPS EOS. 

For $M\la 2~{\rm M}_\odot$, 
the $z_{\rm surf}(M)$ curve for our EOS is quite similar 
to the APR one. 
%
\subsection{Binding energy}
Binding energy of neutron star, $E_{\rm bind}$, is defined as 
the mass defect with respect to a dispersed configuration 
of matter consisting of the same number of baryons, multiplied 
by $c^2$. A dispersed configuration is characterized by 
negligible pressure and negligible gravitational interactions. 
Equivalently, one may define $E_{\rm bind}$ as a net work, 
needed to transform a neutron star into a dispersed configuration 
of matter. In what follows, we will use standard definition of 
$E_{\rm bind}$, i.e., with respect to a dispersed configuration of 
a pressureless cloud of $^{56}{\rm Fe}$ dust, with mass per nucleon 
$m_{\rm Fe}\equiv
 {\rm mass~of} ~^{56}{\rm Fe}~ {\rm atom}/56=
1.6587~10^{-24}~$g. Therefore,
%
\begin{equation}
E_{\rm bind}=\left(A m_{\rm Fe}- M\right)c^2~.
\label{Ebind.eq}
\end{equation}
%
With such a definition, $E_{\rm bind}$ represents a good approximation 
of the binding energy of neutron star with respect to the 
configuration of a presupernova core from which the neutron star 
was formed, via gravitational collapse, as a by-product of the 
type II supernova explosion. Binding energy is plotted versus $M$ in 
Fig. 7. A given $M$, it is somewhat smaller than for a softer FPS EOS. 
However, the maximum value of $E_{\rm bind}$, reached for $M_{\rm max}$, 
is significantly larger for the SLy EOS than for a softer FPS one.
Binding of the neutron star is due to gravitational forces and  
it rises rapidly with $M$. Significantly larger value of $M_{\rm max}$ 
(by 10\%), combined with approximate scaling $E_{\rm bind}\propto 
M^2$ (Lattimer \& Yahil 1989), explain why maximum binding energy 
for our SLy model is some 20\% higher than for the FPS one. 

 The scaling argument is much less precise in the case of comparison 
of maximum $E_{\rm bind}$ for our EOS with the APR one. However, 
let us remind 
that the APR curve above the  asterisk should be treated with 
caution.
 
For measured masses of binary pulsars, we get for our 
 EOS $E_{\rm bind}=
2.3\div 2.7~10^{53}~$erg; corresponding values for the FPS EOS 
are some $2~10^{52}~$erg higher.
\subsection{Moment of inertia}
Most of observed neutron stars are rotating. However, even 
for most rapid millisecond pulsar PSR 1937+21, with rotation 
period $P_{\rm min}^{\rm obs}=1.558$~ ms, 
and angular frequency $\Omega_{\rm max}^{\rm obs}=
2\pi/{\rm period}=
4033$~Hz, rotation implies only small changes of stellar 
structure for  neutron stars with 
$M>1~{\rm M}_\odot$. Therefore, for the 
description of effects of rotation for observed neutron stars 
one can use {\it slow rotation} approximation, in which 
effects of rotation (assumed to be rigid) are treated using 
a lowest order perturbative scheme  (Hartle 1967). 
In this approach, one calculates, in the linear approximation 
in the angular frequency as measured by a distant observer, 
 $\Omega$, total angular momentum of neutron star, 
$J\propto \Omega$ (next order term is cubic in $\Omega$). 
Then, one gets {\it moment of inertia  for slow, rigid rotation} 
as $I=J/\Omega$. Notice that within the slow rotation approximation 
$I$ is independent of $\Omega$ and can be calculated from the 
structure of a non-rotating configuration of neutron star. 
The values of $I$ are plotted, versus $M$, in Fig. 8. 
At given $M$, the value of $I$ for our  EOS 
is significantly higher than for softer FPS EOS. The difference 
rises rapidly with increasing $M$. An even larger difference is noted 
for the maximum value of the 
 moment of inertia, $I_{\rm max}$, reached for 
a mass  slightly lower than $M_{\rm max}$. Indeed, $I_{\rm max}$ 
depends quite sensitively on the stiffness of the EOS of dense matter, 
and this dependence can be approximated by 
$I_{\rm max}/10^{45}~{\rm g~cm^2}\simeq      
(M_{\rm max}/{\rm M}_\odot)(R_{M_{\rm max}}/10~{\rm km})^2$ 
(Haensel 1990). As we noted before, $M_{\rm max}$, and the radius 
at $M_{\rm max}$, denoted by $R_{M_{\rm max}}$, for 
the FPS EOS constitute, respectively, 88\% and 93\% 
of the values obtained for our SLy EOS.The simple 
approximate relation mentioned before, 
derived in (Haensel 1990), implies then that $I_{\rm max}$ 
for the FPS EOS has to be only 76\% of $I_{\rm max}$ obtained with 
our EOS, which nicely reproduces results of exact calculations. 
 The same scaling argument can be applied to ``explain'' 
the difference in $I_{\rm max}$ for our EOS and the APR 
one, in terms of the difference in $M_{\rm max}$ and
 $R_{M_{\rm max}}$.  
\subsection{Effects of rotation at $P \ge 1.558$~ms}
As we mentioned in the preceding subsection, the effect of rotation 
on the structure of neutron stars with $M>1~{\rm M}_\odot$ is 
small, and can be treated as a perturbation. Rigid  rotation 
increases the maximum mass of the SLy neutron stars by less 
than two  percent. The smallness of this effect is readily understood, 
because it is quadratic in 
a dimensionless parameter 
${\overline \Omega}= \Omega/\sqrt{GM/R^3}$ (Hartle 1967). 
At present $P\ge 1.558~$ms, and therefore
${\overline \Omega}^2 \le   
{\overline \Omega}_{\rm max}^{\rm obs}=
0.06$ at $M_{\rm max}$ for the SLy EOS.  
Now, maximal rotation, at $\Omega_{\rm max}\sim \sqrt{GM/R^3}$,  
implies increase of $M_{\rm max}$ by about 20\% (see next 
subsection). Therefore, fractional increase of $M_{\rm max}$ 
connected with ${\overline\Omega}^2\ll 1$ is 
$0.2{\overline\Omega}^2< 2\%$. 
 
However,  
${\overline \Omega}_{\rm max}^{\rm obs}$ increases with decreasing 
neutron star mass, and rotational effects  at 
$P_{\rm min}^{\rm obs}=1.558$~ms 
become decisive for low-mass neutron stars. Namely, low-mass 
neutron stars become significantly flattened and  because of 
strongly 
increasing  radius (with decreasing mass) 
they  approach rapidly the  {\it mass shedding limit}, 
at which gravitational pull at the equator is exactly balanced 
by the centrifugal force. Exact 2-D calculation show that the 
minimum mass of neutron stars rotating rigidly at the 
minimum observed pulsar period of $1.558$~ms is,  
for our  EOS, $M_{\rm min}(1.558~{\rm ms})=
0.61~{\rm M}_\odot$, some seven times higher than for static 
neutron stars (Haensel et al. 2001). 

\begin{figure}
\resizebox{\hsize}{!}{\includegraphics{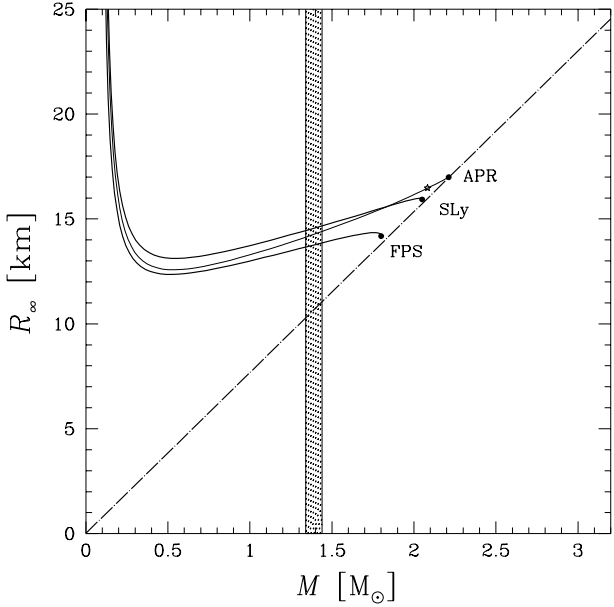}}
\caption{
Apparent radius of neutron star, $R_\infty$, versus gravitational 
mass, $M$, Long-dash-dot straight line corresponds to 
minimum $R_\infty$ at a given $M$ (see the text).
}
 \label{RMinf}
\end{figure}

\subsection{Maximally rotating neutron star models}
Minimum rotation period for  rigidly rotating neutron stars, 
stable with respect to the axisymmetric perturbations, was  
 calculated using the 2-D general relativistic code, based 
on the pseudospectral method for solving partial 
differential equation (Gourgoulhon et al., in preparation). 
We get $P_{\rm min}=0.55~$ms.
For subluminal EOS ($v_{\rm s}\le c$), the values of $P_{\rm min}$ 
can be estimated, with a  good (a few percent) 
precision, using 
an ``empirical formula'' 
$P_{\rm min}/{\rm ms}=0.82\cdot (M_{\rm max}/
{\rm M}_\odot)^{-1/2} 
(R_{M_{\rm max}}/{\rm 10~km})^{3/2}~$, where $M_{\rm max}$ 
and $R_{M_{\rm max}}$ are mass and radius of {\it static} 
configuration with maximum allowable mass (Haensel \& Zdunik 1989, 
Haensel et al. 1995). This empirical formula yields $P_{\rm min}
\simeq 0.57~$ms, which is only 4\% higher than the exact result.  
Using the empirical formula, one gets for the APR EOS the 
value of $P_{\rm min}$ which is a few percent lower than 
for our EOS. 

The maximum mass for rigidly rotating neutron stars 
is $M^{\rm rot}_{\rm max}=2.42~{\rm M}_\odot$, 
 18\% higher than the maximum mass for static 
configurations. For  the FPS EOS, one obtains 
 $M^{\rm rot}_{\rm max}=1.95~{\rm M}_\odot$ (Cook 
et al. 1994), also 18\% higher than for the static configuration. 
Such an increase of $M_{\rm max}$ due to a maximal 
uniform rotation is characteristic of subluminal realistic 
baryonic EOS (Lasota et al. 1996).  
\section{Neutron star models vs.  observations}
\subsection{Masses}
The value of $M_{\rm max}=2.05~{\rm M}_\odot$ is sufficiently 
high to be consistent with measured masses of binary radio pulsars 
and with estimates of masses of X-ray pulsars and X-ray bursters. 
It is a little too low (by a few percent) to explain the 
upper-peak frequency $1.06\pm 0.02$~kHz of the QPOs in 
4U 1820-30 via the orbital motion of plasma clumps in the 
marginally stable circular orbit, which would require 
$M>2.1~{\rm M}_\odot$. However, the last constraint may be 
not valid because of doubts  connected with interpretation 
of the kHz QPOs in this and other LMXBs. These doubts result  
from  observed  changes  of the difference between the upper and 
lower peak frequencies, which within the beat-frequency model 
(Miller et al. 1998 and references therein) is identified with 
rotation frequency of neutron star, and should therefore be 
constant, apart from slight increase due to the accretion 
spin-up.
\subsection{Apparent radii of  isolated closeby neutron stars}
Measuring the spectrum of photons emitted from the surface of a 
solitary neutron star, combined with knowledge of distance 
(from the annual parallax) enables one, in principle, to 
determine total photon luminosity, effective surface temperature 
and  {\it apparent radius} of neutron star. 
Recently, such studies have been carried out for Geminga 
(Golden \& Shearer 1999) and 
RX J185635-3754 
(Walter 2001, Pons et al. 2001). 

The apparent (or radiation) radius, 
measured by a distant observer, $R_\infty$, is related to 
$R$ by
\begin{equation}
R_\infty={R\over \sqrt{1-r_{\rm g}/R}}= (1+z_{\rm surf})R~. 
\label{Rinf}
\end{equation}
The values of $R_\infty$ versus $M$ are plotted in Fig.\ref{RMinf}. 
The $R_\infty(M)$ plot is very different from the $R(M)$ one (Haensel 2001). 
In particular, $R_\infty(M_{\rm max})$  {\bf is not} the minimum value of 
$R_\infty$; the minimum  is usually reached at $M\sim 0.5~{\rm M}_\odot$, 
and for our EOS is 13.1 km, to be compared with slightly smaller minimum 
values  of 12.3 km and 12.5 km for the FPS and APR EOS. 

A strict lower bound on $R_\infty$ at a given $M$ results from the very 
definition of $R_\infty$, and does not depend on any physical constraint:
 $R_\infty(M)>R_{\rm \infty,min}=7.66~(M/{\rm M}_\odot)$~km (Lattimer 
\& Prakash 2001, see also Haensel 2001). 
As one sees in Fig.\ref{RMinf}, the values of $R_\infty$ 
at $M_{\rm max}$ are extremely close (within less than 1\%) 
to $R_{\rm \infty,min}$. This 
property has been explained  in (Haensel 2001).

Central value of $R_\infty$ for Geminga, obtained by Golden \& Shearer (1999) 
using the best-fit 
model atmosphere spectra, cannot be explained by our EOS, and actually - {\it 
by none} of existing baryonic EOS of dense matter (it could be modeled 
by a small mass strange star covered by a normal matter layer, to produce 
the observed photon spectrum). However, the uncertainty in the extracted 
value of $R_\infty$ is large: it stems mainly from the uncertainty in the 
distance to Geminga (assumed to be $d=159$~pc), but a poor knowledge 
of the photon spectrum plays also an important role. One therefore might 
argue, that because of these uncertainties the measured value of $R_\infty$ 
cannot exclude  our and other baryonic EOS of dense matter at a reasonably 
high confidence level of 95\%.  

In the case of 
RX J185635-3754 
contradiction between extracted value of $R_\infty$, and theoretical models of 
neutron stars based on our EOS (and on other available models of dense 
 matter) is even more dramatic (Pons et al. 2001).
\footnote{Low-mass strange quark stars covered with a thin normal matter envelope 
are excluded too because the best-fit redshift $z_{\rm surf}
\simeq 0.3-0.4$ (Pons et al. 2001).}
The central best-fit value of $R_\infty$ is 8.2 km (Fe atmosphere) and 
7.8 km (Si-ash atmosphere), at assumed distance  $d=61$~pc.
Non-uniformity of surface temperature, consistent with observational 
constraints, 
does not allow to remove this conflict between theory and observations. 
Unfortunately, proper inclusion of effects of surface magnetic field 
is not possible because of non-availability of magnetized heavy-metal 
atmosphere models (Pons et al. 2001). 
One may only hope, that the problem 
of the conflict between theoretical and measured $R_\infty$ of closeby 
isolated neutron stars will be solved in the future studies. 

Very recently, Rutledge et al. (2001) proposed a method of measuring 
$R_\infty$ of neutron stars, observed as X-transients in globular 
clusters. They studied transient X-ray source CXOU 132619.7-472910.8 
in NGC 5139. Fitting its photon spectrum with H-atmosphere model, 
they obtained, at 90\%  confidence level, $R_\infty=14.3 \pm 
2.5$~km, which is  consistent with our EOS, and with FPS, APR 
and many other available EOS of dense matter. 
 This method of measuring $R_\infty$ 
seems to be very promising, because 	both distance and interstellar hydrogen 
column density are relatively well known for globular clusters. 
 
\subsection{Surface redshift}
As of this writing (June 2001), 
 the only reliable evaluation of $z_{\rm surf}$ seems to be 
connected with extraordinary gamma-ray burst GRB 790305b 
(of March 5th, 1979) from the soft-gamma repeater SGR 0526-66 
associated with supernova remnant N49 in Large Magellanic Cloud. 
The spectrum of this gamma-ray burst exhibited a prominent 
emission line at $430\pm30$ keV, with full width at half-minimum 
$\simeq 150$~keV (Mazets et al. 1981, 1982). Assuming that the line 
originated from $e^+e^-\longrightarrow 2\gamma$, and that line 
broadening resulted from the thermal motion in 
the plasma, one gets, after taking due account of the thermal 
blueshift (see, e.g., Higdon \& Lingenfelder 1990), 
$z_{\rm surf}=0.23\pm 0.07$. As one sees in Fig.\ref{zM}, 
such surface redshift is predicted for neutron stars 
of $M=1\div 1.6~{\rm M}_\odot$, while the central measured value 
of 0.23  corresponds for our EOS to a  
neutron star with  $M\simeq 1.4~{\rm M}_\odot$.  
\subsection{Binding energy and SN 1987A}
 The appearence  of SN1987A in the UV and optical domain 
   was preceded by a burst of neutrinos, detected 
on the Earth by  neutrino detectors. 
The total of 25 events of the 
absorption of ${\overline\nu}_e$ 
on protons were  
registered within $\sim 10$ s. Analysis of these events,  combined 
with:  knowledge of detectors 
properties,  assumption of spherical symmetry, knowledge of the 
distance to Large Magellanic Cloud, and basic features of the 
SN II theory, enabled the evaluation of the 
total energy of the neutrino burst as $E_\nu\simeq (2.5\pm 1)  
~10^{53}~$erg (Lattimer \& Yahil 1989). As about $99\%$ of 
the total energy release in a SN II explosion is emitted 
in a neutrino burst, this was actually the measurement of the binding 
energy of a newly born neutron star, $E_{\rm bind}\simeq E_\nu$. 
Stellar evolution theory tells us, that the neutron star born in 
SN1987A had $M=1.2\div 1.5~{\rm M}_\odot$. This restricts the 
area in the $E_{\rm bind}-M$ plane in Fig.\ref{EbM} to a small shaded 
rectangle. This rectangle is nicely consistent with binding energies 
of neutron stars predicted by our EOS.  
\subsection{Crustal moment of inertia and pulsar glitches}
It is widely   believed that sudden spin ups of radio pulsars, 
called {\it glitches}, are due to the angular momentum transfer 
from a specific, weakly coupled to the rest of the star,  
 superfluid component (dripped neutrons in the inner crust) -  
to the rest of the neutron star body (lattice of nuclei in the crust 
plus the liquid core) (Alpar et al. 1984). 
 Since the discovery of glitches in 1969, more 
than thirty of them have been observed. Particularly large number 
of glitches have been detected in the 
timing of the Vela pulsar (thirteen during 
the period 1969-1999). The set of data on the Vela glitches 
was used by Link et al. (1999) to derive a constraint on the 
neutron star crustal moment of inertia, $I_{\rm crust}/I>1.4\%$. 
This constraint is satisfied  by our EOS, provided the neutron star 
 mass
is below $1.75~{\rm M}_\odot$. 
  Let 
us mention that in the case of a softer FPS EOS 
 neutron star mass should not  
not exceed $1.6~{\rm M}_\odot$ in order to satisfy this 
 constraint. 
\subsection{Neutron star cooling}
For our specific SLy model, the threshold density for the 
direct electron Urca process, $1.35~{\rm fm^{-3}}$, 
 is above the maximum central 
density of neutron stars. Taken at its face value, this 
would mean that the SLy model does not allow for the 
direct Urca process involving nucleons in neutron 
stars. One should keep in mind however, that the specific 
SLy4 model that we used is only one of a larger  set of the 
SLy family. We remind that for some versions of these 
effective nuclear interactions direct Urca process was 
allowed in massive neutron stars (see Chabanat et al. 1997). 
Moreover, the SLy forces were constructed to reproduce 
best variational calculations for high density  pure 
{\it neutron matter} with realistic neutron-neutron 
potentials, and therefore proton fraction  they yield 
should not be considered at the same footing as 
the  EOS for neutron star matter, in which 
protons play a rather small role. 
\section{Summary and conclusion}
We calculated the EOS of neutron star matter, which describes in 
a physically unified way both the crust and the liquid core. 
The EOS, valid from  $10^8~{\rm g~cm^{-3}}$ up to the 
maximum density reachable within neutron stars, 
 was  based on the recently derived 
SLy effective nucleon-nucleon interaction, which,  due to 
its construction method,   is particularly 
suitable for the description of strong interactions in the nucleon 
component of dense neutron star matter. 
Calculations were done assuming ground state of neutron star 
crust, and the ``minimal'' $npe\mu$ composition of the 
liquid core.  
The minimum and maximum mass of non-rotating 
neutron stars are $0.09~{\rm M}_\odot$ and $2.05~{\rm M}_\odot$ 
respectively. Rigid rotation at the minimum observed 
pulsar period 1.558 ms increases the maximum mass by only 
about 1\%, but  effect on the minimum mass is large: 
it increases  up to $0.61~{\rm M}_\odot$.  

Our model of matter at supranuclear densities is the 
simplest possible, and is based on experimental nuclear 
physics  and relatively precise  many-body calculations 
of dense neutron matter. We did not  consider possible  
dense  matter constituents, for which strong interactions are 
poorly known (hyperons), or which are hypothetical 
(pion and kaon condensates, quark matter). 
Such a model as that proposed in the present paper
may seem very simple - as  compared to a rich spectrum 
of possibilities considered in the literature on 
the constitution of dense neutron star cores. 
However, it has virtue of giving unified 
description of all interior of neutron star, 
and is firmly based  on the most solid sector 
of our knowledge of nuclear interactions. 
%
\begin{acknowledgements}
We express our gratitude to A. Potekhin for reading the 
manuscript, and for  remarks and comments which helped to improve 
the present paper.  We are also grateful to him  for his precious help in 
the preparation of figures.  
This research
was partially supported by the KBN grant No. 5P03D.020.20 and 
by the CNRS/PAN Jumelage Astrophysique Program. 
\end{acknowledgements}

\end{document}